\documentclass[12pt]{article}

\usepackage{amsfonts}
\usepackage{amsmath}
\usepackage{esint}
\usepackage[margin=1in]{geometry}
\usepackage{graphicx}
\usepackage{epstopdf}
\usepackage{amssymb}
\usepackage{cite}
\usepackage{multicol}
\usepackage{color}
\usepackage[section]{placeins}
\usepackage[linesnumbered,ruled,vlined]{algorithm2e}
\usepackage{framed}
\usepackage[font=small,labelfont=bf]{caption}
\usepackage{hyperref}
\usepackage{orcidlink}

\hypersetup{
  colorlinks = true, %
  urlcolor   = black, %
  linkcolor  = black, %
  citecolor  = black %
}

\SetKwInput{KwInput}{Input}      
\SetKwInput{KwOutput}{Output}   
\SetKwComment{Comment}{/* }{ */}

\newcommand{\ds} {\displaystyle}
\newcommand{\Frac}[2]{\ds \frac{#1}{#2}}

\newcommand{\R}{{\mathbb R}}

\newcommand{\mbf}[1]{\mathbf{#1}}
\newcommand{\bsym}[1]{\boldsymbol{#1}}

\title{A Systematic Particle Filter for Estimating Time-Varying Parameters in Advection-Diffusion Equations with Source Terms}
\author{Andrea Arnold~{\orcidlink{0000-0003-3003-882X}}}
\date{}

\begin{document}
\maketitle

\small 
\centerline{Department of Mathematical Sciences, Worcester Polytechnic Institute, Worcester, MA, USA}
\vspace{.2cm}

\centerline{E-mail: anarnold@wpi.edu}

\normalsize

\bigskip

\begin{abstract}
\noindent
Many real-world systems modeled using partial differential equations (PDEs) involve unknown parameters that must be estimated from limited, noisy system observations.
While typically assumed to be constants, some of these unobserved parameters may vary with time.
This work proposes a two-phase, offline-online numerical procedure for systematically estimating and quantifying uncertainty in time-varying parameters (TVPs) in time-dependent PDEs, specifically focusing on advection-diffusion models with TVPs involved in the source terms.
Numerical results on a set of one-dimensional test problems demonstrate the effectiveness of the proposed estimation procedure in tracking unknown TVPs of different forms, while simultaneously estimating variance parameters affecting the TVP evolution model, from partial, noisy observations of the solution at discrete spatial locations and times.\\

\end{abstract}


\section{Introduction}

Many real-world systems modeled using differential equations involve unknown parameters that must be estimated from limited, noisy system observations.
While these parameters tend to be modeled as constants, certain systems may involve unmeasured parameters that vary with time.
In the setting of partial differential equations (PDEs), time-varying parameters (TVPs) may arise, e.g., in modeling thermal fields \cite{Wang2020}, the pricing of stock options \cite{Rodrigo2006}, transport in porous media \cite{Selvadurai2004}, and laser-tissue interactions \cite{ArnoldFichera2022}.
It is therefore of interest to estimate and quantify uncertainty in unknown TVPs in order to make data-informed forecast predictions with PDE models.

This work proposes a two-phase, offline-online numerical procedure for systematically estimating and quantifying uncertainty in TVPs in time-dependent PDEs. 
In particular, we focus on advection-diffusion equations with time- and spatially-dependent source terms, where the source terms involve a TVP of interest. 
Under the proposed framework, the PDE spatial variables are discretized offline via a method of lines (MOL) approach, forming a system of coupled ordinary differential equations (ODEs).
The online phase leverages the work in \cite{Arnold2023}, which developed a systematic particle filtering algorithm for estimating TVPs in ODE systems using a hierarchical Bayesian modeling framework.

Various approaches have been used to estimate constant parameters in PDE models, many inspired from comparable approaches for ODEs.
These include use of artificial neural networks \cite{Jamili2021}, multiple shooting techniques \cite{Muller2002}, derivative-free optimization \cite{Carvalho2015}, residual-based recursive methods \cite{Guo2009}, Gaussian processes \cite{Raissi2017, Rai2019}, and parameter cascading \cite{Xun2013}. 
Several works have extended these techniques to estimate spatially-varying parameters; e.g., \cite{Abdolee2014, Kramer2013, Young2012}.
Considering the use of particle filter (PF) or sequential Monte Carlo (SMC) algorithms in this context, the work in \cite{Beskos2015} uses an SMC method for Bayesian inference of elliptic PDEs (estimating constants) while the work in \cite{Kantas2014} applies an SMC approach to estimate the initial condition of a two-dimensional (2D) Navier-Stokes equation.

Few works in the literature have specifically addressed TVP estimation in the setting of PDEs.
The work in \cite{Wang2020} utilizes B-spline basis functions to form a TVP approximation model, estimating the constant coefficients of the spline representation in a three-dimensional (3D) spatio-temporal thermal field.
Similarly, the work in \cite{Zhang2017} applies parameter cascading to estimate the constant coefficients of a basis function representation of varying-coefficients in PDEs.
The work in \cite{Qin2021} provides a data-driven method based on deep neural networks (DNNs) to learn nonautonomous systems with time-dependent inputs, including one example using a 1D heat equation with a time-dependent parameter in the source term; however, the goal in this work was not to explicitly estimate the TVP, instead using quadratic polynomial interpolation to approximate the TVP within the DNN framework to predict the system dynamics.

More closely related to the proposed methodology, the work in \cite{ArnoldFichera2022} directly addresses TVP estimation using an ensemble Kalman filter-based approach to estimate time-varying absorption and scattering coefficients in a 3D heat equation with source modeling laser-tissue interactions.   
Recent work in \cite{Lin2025} applies a source term estimation technique using unsteady adjoint equations combined with nonsequential Bayesian inference to estimate time-dependent, localized sources in an advection-diffusion model for dispersing pollutants (e.g., chemical leak) around buildings.

\paragraph{Paper Contributions and Organization}

This paper contributes a new approach to solving TVP estimation inverse problems in applications involving time-dependent PDE models, offering a method for systematic estimation and uncertainty quantification under a sequential Bayesian inference framework.
The proposed two-phase, offline-online scheme allows for direct estimation of the PDE model TVPs while simultaneously tracking nuisance parameters affecting the TVP evolution model. 
We demonstrate the effectiveness of this approach on advection-diffusion-type models, where the TVP of interest lies in the additive source term.

The remainder of the paper is organized as follows:
Section~\ref{Sec:Methods} gives an overview of the TVP estimation inverse problem and details the proposed two-phase estimation procedure.
Section~\ref{Sec:Results} applies the proposed methodology to numerical examples using two different PDE test models.
Section~\ref{Sec:Discussion} discusses the main take-aways and concludes the paper.


\section{Methods}
\label{Sec:Methods}

This section describes the TVP estimation inverse problem of interest in this work, then details the two-phase, offline-online numerical procedure for TVP estimation and uncertainty quantification. 

\subsection{Problem Setup}
In setting up the inverse problem, consider a $q$-dimensional, time-dependent PDE of the general form
\begin{equation}
\frac{\partial u}{\partial t} = \mathcal{F}\big( t, \mbf{x}, u, \frac{\partial u}{\partial \mbf{x}_i}, \frac{\partial^2 u}{\partial \mbf{x}_i^2}, \frac{\partial^2 u}{\partial \mbf{x}_i \partial \mbf{x}_j}, \dots; \bsym{\theta}(t) \big) 
\end{equation}
with solution $u = u(\mbf{x},t)$, where $\mbf{x} = (x_1,\dots, x_q)$ denotes the spatial variables (and typically $q = 1$, 2, or 3), $t$ represents the time variable, and $\bsym{\theta}(t)$ is a vector of time-dependent model parameters.
While some of the model parameters may be constant in time, here we assume more generally (and emphasize with the notation) that the parameters of interest are time-varying.
We further assume prescribed boundary and initial conditions as relevant for the particular PDE model at hand. 
Given partial, noisy observations of the model states (i.e., corrupt observations of the system at a subset of spatial locations) at discrete observation times, 
the inverse problem we aim to address is to estimate and quantify uncertainty in the unknown TVP vector $\bsym{\theta}(t)$ as well as the solution $u(\mbf{x},t)$ at some discrete times and spatial locations.

For the problems in this work, we will focus on advection-diffusion PDEs of the form
\begin{equation}\label{eq:full_advec}
\frac{\partial u}{\partial t} + \mbf{v} \cdot \nabla u = \alpha \nabla^2 u + S(\mbf{x},t)
\end{equation}
where $\nabla$ and $\nabla^2$ denote the gradient and Laplacian operators, respectively, $\mbf{v}$ is the velocity field, $\alpha$ is the diffusion coefficient, and $S(\mbf{x},t)$ is a source term that may depend on space and/or time.
Advection-diffusion equations are commonly used to model the transport of quantities of interest (e.g., heat, particles, chemical concentrations) through various media \cite{Selvadurai2004, Beard2000, Luce2013, Koch1987}. 
We further assume that the TVPs of interest appear in the PDE model source term, i.e., that $S(\mbf{x},t) = S(\mbf{x},t; \bsym{\theta}(t))$, and that the other model terms have fixed coefficients. 
Time-varying source term parameters may appear in applications modeling, e.g., laser-tissue interactions with time-varying tissue absorption \cite{ArnoldFichera2022} or the release of a pollutant from point sources with time-dependent intensity \cite{Lin2025}.

\subsection{Two-Phase Estimation Procedure}

The proposed systematic approach for estimating and quantifying uncertainty in the unknown TVPs of interest relies on a two-phase, offline-online procedure: the offline phase utilizes an MOL discretization in the spatial variables of the PDE model, resulting in a system of coupled ODEs, while the online phase employs a systematic particle filtering algorithm to track the quantities of interest (i.e., the ODE model state variables, TVPs, and corresponding TVP evolution model variance parameters).  
We detail each phase in the sections that follow.

\subsubsection{Offline Phase: Spatial Discretization}

During the offline phase, we first discretize the spatial variables $\mbf{x}$ in \eqref{eq:full_advec} over a finite domain in each variable to form a mesh of discrete spatial locations, then approximate the spatial derivatives to form a continuous-time system of coupled ODEs in an MOL approach.
For example, considering one spatial variable $x\in[0,L]$, we implement a discretization $x_i$, $i=0,\dots, M$, with equidistant spatial locations $h=\Delta x$ units apart. 
This approach can be extended to form a mesh of equispaced points over an area (in 2D) or volume (in 3D).

We then apply finite difference schemes to approximate the spatial derivatives in the gradient and Laplacian terms in \eqref{eq:full_advec}; e.g., the central difference formulas for approximating first and second derivatives in one variable yield 
\begin{equation}\label{eq:FD1}
\frac{\partial u(x_i,t)}{\partial x}  \approx \frac{u(x_i+h,t)-u(x_i-h,t)}{2h} = \frac{u(x_{i+1},t)-u(x_{i-1},t)}{2h} 
\end{equation}
and
\begin{equation}\label{eq:FD2}
\frac{\partial^2 u(x_i,t)}{\partial x^2} \approx \frac{u(x_i+h,t)-2u(x_i,t)+u(x_i-h,t)}{h^2} = \frac{u(x_{i+1},t)-2u(x_i,t)+u(x_{i-1},t)}{h^2} , 
\end{equation}
respectively. 
Applying finite difference approximations to the spatial derivatives in \eqref{eq:full_advec} and solving for the time derivative at each spatial mesh point, stacking the spatial locations into a single vector, results in a system of coupled ODEs of the form
\begin{equation}\label{eq:ODEs}
\frac{d\mbf{u}(t)}{dt} = f(t,\mbf{u}(t),\bsym{\theta}(t)), 
\end{equation}
where $\mbf{u}=\mbf{u}(t)\in\R^{d}$ denotes the states of the ODE model (i.e., the continuous-time functions of the PDE solution at the discretized spatial locations), $d$ depends on the number of spatial mesh points and on the boundary conditions of the PDE model, and the initial value $\mbf{u}(0)$ is set based on the given initial condition $u(\mbf{x},0)$ of the PDE.   
While here we focus on finite difference approximation on an equispaced mesh, we note that, in general, the mesh points need not be equispaced, and we could alternatively use a finite element discretization, as considered, e.g., in \cite{Pacheco2025}.

\subsubsection{Online Phase: TVP Estimation via Particle Filtering}
\label{Sec:PF}

Once formulated as system of ODEs as in \eqref{eq:ODEs}, we apply a sequential Bayesian inference procedure to estimate and quantify uncertainty in the quantities of interest, namely, the TVPs as well as the ODE model states (representing PDE solution trajectories at the discretized spatial locations). 
To this end, we leverage a systematic particle filtering algorithm (PF-TVP+) designed for TVP estimation inverse problems; this algorithm was proposed in \cite{Arnold2023} to address TVP estimation in dynamical systems governed by ODEs.
In this work, we extend the approach to solve inverse problems where the forward dynamics are driven by coupled ODE systems representative of time-dependent PDEs.
Here we outline the steps of the PF-TVP+ algorithm as applied to the problem at hand; for more details on the statistical derivation, see \cite{Arnold2023}.
For sake of notation, in the description that follows, we let $\mbf{u}_{\cdot,j}$ denote all components of the state vector $\mbf{u}(t)\in\R^{d}$ (or its stochastic representation) as approximated at time $t_j$.

We begin by defining stochastic processes $\{ U_j \}_{j=0}^\infty$, $\{ \Theta_j \}_{j=0}^\infty$, and $\{ Y_j \}_{j=1}^\infty$, with random variables representative of the model states, TVPs, and observations, respectively.  
Assuming these processes adhere to the Markov property, we can write the following state-space model:
\begin{eqnarray}
U_{j+1} &=& F(U_j, \Theta_j) + V_{j+1}, \quad V_{j+1} \sim \mathcal{N}(\mbf{0}, \mathsf{C}_{j+1}) \label{eq:state_ev}  \\[0.2cm]
\Theta_{j+1} &=& \Theta_j + E_{j+1}, \quad E_{j+1} \sim \mathcal{N}(\mbf{0}, \mathsf{E}_{j+1}) \label{eq:TVP_ev}  \\[0.2cm]
Y_{j+1} &=& G(U_{j+1}) + W_{j+1}, \quad W_{j+1} \sim \mathcal{N}(\mbf{0}, \mathsf{D}_{j+1}) \label{eq:obs_ev}  
\end{eqnarray}
where the state evolution in \eqref{eq:state_ev} is driven by the numerical solution to the ODE system in \eqref{eq:ODEs} from time $t_j$ to $t_{j+1}$ plus innovation, 
the TVP evolution in \eqref{eq:TVP_ev} is modeled as a random walk process, and 
the observation equation in \eqref{eq:obs_ev} prescribes the likelihood function.
Here we model the state, parameter, and observation noise processes $\{ V_j \}_{j=0}^\infty$, $\{ E_j \}_{j=0}^\infty$, and $\{ W_j \}_{j=0}^\infty$, respectively, as following Gaussian distributions with zero mean and prescribed covariance matrices $\mathsf{C}_{j+1}$, $\mathsf{E}_{j+1}$, and $\mathsf{D}_{j+1}$.
The parameter noise covariance (sometimes referred to as the parameter drift covariance) $\mathsf{E}_{j+1}$ is of particular interest in this work, since it has a significant effect on the resulting TVP estimates \cite{Arnold2023}.
We therefore design a particle filtering scheme to systematically estimate the relevant entries of $\mathsf{E}_{j+1}$ along with the TVPs, under the simplifying assumptions that $\mathsf{E}_{j+1}$ is a diagonal, time-invariant matrix, such that $\mathsf{E}_{j+1} = \mathsf{E} = \text{diag}(\bsym{\sigma}_\mathsf{E}^2)$ and we want to learn $\bsym{\sigma}_{\mathsf{E}}^2$.

Starting at time index $j=0$, we can use Bayes' Theorem to formulate a sampling scheme to perform the sequential updates
\begin{equation}
\pi(\mbf{u}_{\cdot, j}, \bsym{\theta}_j, \bsym{\sigma}_{\mathsf{E}, j} | \mbf{y}_j) \longrightarrow \pi(\mbf{u}_{\cdot, j+1}, \bsym{\theta}_{j+1}, \bsym{\sigma}_{\mathsf{E}, j+1} | \mbf{y}_{j+1}), \quad j = 0, \dots, J-1,
\end{equation}
where the posterior probability distribution $\pi(\mbf{u}_{\cdot, j+1}, \bsym{\theta}_{j+1}, \bsym{\sigma}_{\mathsf{E}, j+1} | \mbf{y}_{j+1})$ of the model states, TVPs, and parameter drift standard deviation parameters (referred to as parameter drift coefficients moving forward) conditioned on the data at time $t_{j+1}$ becomes the prior distribution for the next time step.
The algorithm proceeds as follows.

Given a joint prior distribution $\pi(\mbf{u}_{\cdot,0},\bsym{\theta}_0,\bsym{\sigma}_{\mathsf{E},0}|D_0)$, where $D_0 = \emptyset$ denotes the data at time $t_0$, we draw an initial sample of state, TVP, and parameter drift coefficient particles and assign equal weights:
\begin{equation}
\mathcal{S}_0 = \big\{ \big(\mbf{u}_{\cdot,0}^{(n)}, \bsym{\theta}_0^{(n)}, \bsym{\sigma}_{\mathsf{E},0}^{(n)}, w_0^{(n)}\big)\big\}_{n=1}^N, \quad w_0^{(1)} = \cdots = w_0^{(N)} = \frac{1}{N}.
\end{equation}
The initial sample mean and covariance of the parameter drift coefficients are given by
\begin{equation}
\bsym{\bar{\sigma}}_{\mathsf{E},0} = \sum_{n=1}^N w_0^{(n)} \bsym{\sigma}_{\mathsf{E},0}^{(n)} \quad \text{and} \quad \mathsf{S}_0 = \sum_{n=1}^N w_0^{(n)} \big(\bsym{\sigma}_{\mathsf{E},0}^{(n)}-\bsym{\bar{\sigma}}_{\mathsf{E},0}\big) \big(\bsym{\sigma}_{\mathsf{E},0}^{(n)}-\bsym{\bar{\sigma}}_{\mathsf{E},0}\big)^{T}.
\end{equation}
Setting the time index $j=0$, we cycle through the following steps while $j<J$, where $J$ denotes the number of observation times:
\begin{enumerate}
\item Shrink the parameter drift coefficient particles toward the sample mean:
\[
\widehat{\bsym{\sigma}}_{\mathsf{E},j}^{(n)} = a\bsym{\sigma}_{\mathsf{E}, j}^{(n)} + (1-a)\bsym{\bar{\sigma}}_{\mathsf{E},j},  
\]
where the shrinking factor $a = (3\delta-1)/(2\delta)$ and discount factor $1/3 < \delta < 1$ are set to avoid information loss in estimating these parameters, as detailed in \cite{LiuWest2001, West1993a, West1993b}. 
\item Propagate the state particles forward in time to generate the state predictor particles: 
\[
\widehat{\mbf{u}}_{\cdot, j+1}^{(n)} = F\big(\mbf{u}_{\cdot, j}^{(n)}, \bsym{\theta}_j^{(n)}\big), \quad n = 1, \dots, N.
\]
Here the forward operator $F$ represents the numerical solution of the ODE system in \eqref{eq:ODEs} from time $j$ to $j+1$ using an appropriate solver (e.g., to account for potential stiffness). 
\item Given the observed data $\mbf{y}_{j+1}$, use the likelihood function to calculate and normalize the particle fitness weights:
\[
g_{j+1}^{(n)} = w_j^{(n)}  \pi(\mbf{y}_{j+1} | \widehat{\mbf{u}}_{\cdot, j+1}^{(n)}) , \quad g_{j+1}^{(n)} \ \leftarrow \ \frac{g_{j+1}^{(n)} }{\sum_n g_{j+1}^{(n)}} 
\]
\item Use the fitness probabilities to draw (with replacement) the auxiliary indices $\ell_n \in \{1, 2, \dots, N\}$, $n=1,\dots, N$:
\[
P\{\ell_n = k \} = g_{j+1}^{(k)}.
\]
\item Reshuffle the state, TVP, parameter drift coefficient, and state predictor samples using the auxiliary indices:
\[
\mbf{u}_{\cdot, j}^{(n)} \ \leftarrow \ \mbf{u}_{\cdot, j}^{(\ell_n)} , \quad \bsym{\theta}_j^{(n)} \ \leftarrow \ \bsym{\theta}_j^{(\ell_n)} , \quad \widehat{\bsym{\sigma}}_{\mathsf{E}, j}^{(n)} \ \leftarrow \ \widehat{\bsym{\sigma}}_{\mathsf{E}, j}^{(\ell_n)} , \quad \widehat{\mbf{u}}_{\cdot, j+1}^{(n)} \ \leftarrow \ \widehat{\mbf{u}}_{\cdot, j+1}^{(\ell_n)} , \quad n = 1, \dots, N.
\]
\item Innovate the reshuffled state particles:
\[
\mbf{u}_{\cdot, j+1}^{(n)} = \widehat{\mbf{u}}_{\cdot, j+1}^{(n)} + \mbf{v}_{j+1}^{(n)}, \quad \mbf{v}_{j+1}^{(n)}\sim\mathcal{N}(\mbf{0},\mathsf{C}_{j+1}), \quad n = 1, \dots, N, 
\]
where each $\mbf{v}_{j+1}^{(n)}$ is a realization of the state evolution noise $V_{j+1}$ in \eqref{eq:state_ev}.
\item Evolve the reshuffled parameter drift coefficient particles:
\[
\bsym{\sigma}_{\mathsf{E}, j+1}^{(n)} = \widehat{\bsym{\sigma}}_{\mathsf{E}, j}^{(n)} + \bsym{\zeta}_{j+1}^{(n)}, \quad  \bsym{\zeta}_{j+1}^{(n)}\sim\mathcal{N}(\mbf{0},r^2\mathsf{S}_j), \quad n = 1, \dots, N,
\]
where $r^2 = 1-a^2$ and we note that this parameter evolution is artificial, since we assume $\bsym{\sigma}_{\mathsf{E}}$ is time-invariant.
\item Form the parameter drift covariance matrices:
\[
\mathsf{E}_{j+1}^{(n)} = \text{diag}\big((\bsym{\sigma}_{\mathsf{E}, j+1}^{(n)})^2\big), \quad n = 1, \dots, N.
\]
\item Propagate the reshuffled TVP particles:
\[
\bsym{\theta}_{j+1}^{(n)} = \bsym{\theta}_j^{(n)} + \bsym{\xi}_{j+1}^{(n)}, \quad \bsym{\xi}_{j+1}^{(n)}\sim\mathcal{N}(\mbf{0},\mathsf{E}_{j+1}^{(n)}), \quad n = 1, \dots, N,
\]
where each $\bsym{\xi}_{j+1}^{(n)}$ is a realization of the parameter evolution noise $E_{j+1}$ in \eqref{eq:TVP_ev}.
\item Using the observed data $\mbf{y}_{j+1}$, recalculate and normalize the weights for each $n$:
\[
w_{j+1}^{(n)} = \Frac{ \pi(\mbf{y}_{j+1} | \mbf{u}_{\cdot, j+1}^{(n)} ) }{\pi(\mbf{y}_{j+1} | \widehat{\mbf{u}}_{\cdot, j+1}^{(n)}) }, \quad w_{j+1}^{(n)} \ \leftarrow \ \frac{w_{j+1}^{(n)} }{\sum_n w_{j+1}^{(n)}}.
\]
\item Recalculate the mean and covariance of the drift coefficient sample:
\[
\bsym{\bar{\sigma}}_{\mathsf{E}, j+1} = \sum_{n=1}^N w_{j+1}^{(n)} \bsym{\sigma}_{\mathsf{E}, j+1}^{(n)}, \quad \mathsf{S}_{j+1} = \sum_{n=1}^N w_{j+1}^{(n)} \big(\bsym{\sigma}_{\mathsf{E}, j+1}^{(n)}-\bsym{\bar{\sigma}}_{\mathsf{E}, j+1}\big) \big(\bsym{\sigma}_{\mathsf{E}, j+1}^{(n)}-\bsym{\bar{\sigma}}_{\mathsf{E}, j+1}\big)^{T}.
\]
\item Set $j \leftarrow j+1$, and repeat from Step 1 (until $j = J$).
\end{enumerate}
Note that, at each time step, we can use the current particles to compute the sample mean (referred to as the PF mean moving forward) and calculate the corresponding Bayesian credible intervals, which provide a natural measure of uncertainty in the mean estimates (with wider credible intervals indicating less certainty).


\section{Numerical Results}
\label{Sec:Results}

We demonstrate the effectiveness of the proposed TVP estimation approach with numerical examples computed using two different one-dimensional (1D) advection-diffusion-type PDE models:
(i) a 1D advection equation with a logistic source term, and (ii) a 1D heat equation with a Gaussian source term that has time-varying amplitude.
Recall that the 1D advection-diffusion equation with source term takes the general form
\begin{equation}
\frac{\partial u}{\partial t} + v \frac{\partial u}{\partial x} = \alpha \frac{\partial^2 u}{\partial x^2} + S(x,t), \quad x\in[0,L], \quad t\geq0,
\end{equation}
where $u=u(x,t)$ is the solution, we assume that the source term $S(x,t) = S(x,t; \theta(t))$ depends on the TVP of interest, and we further assume finite spatial and time domains for sake of numerical simulation.
Here the velocity $v$ and diffusion coefficient $\alpha$ are assumed to be known, fixed constants, and our goal is to estimate $\theta(t)$ in the source term for each example, while simultaneously tracking the corresponding TVP drift coefficient, i.e., $\sigma_\mathsf{E}$ as defined in Section~\ref{Sec:PF}.

For each of the examples that follow, we assume partial, noisy observations of the solution $u$ at a subset of the discretized spatial locations $x_i$ and discrete times $t_j$; i.e.,
\begin{equation}
\mbf{y}_j = g(\mbf{u}^*(t_j)) + \bsym{\varepsilon}_j, \quad t_1 < t_2 < \cdots < t_J,
\end{equation}
where $\mbf{u}^*(t) = (u(x_0,t),\dots,u(x_M,t))$ denotes the true solution at the spatial locations $x_i$, $i = 0,\dots,M$, $g:\R^{M+1}\times\R\rightarrow\R^m$, $m < M+1$, is the observation mapping, and each $\bsym{\varepsilon}_j \sim \mathcal{N}(\mbf{0}, \sigma^2_\text{noise}\mathsf{I}_m)$ is a realization of Gaussian random noise with standard deviation $\sigma_\text{noise}$ set to be approximately 20\% the average standard deviation of the true solution trajectories.
For the online phase in each case, we use a sample size of $N=1,000$ particles, 
draw the initial state and TVP samples from uniform prior distributions with ranges set to be 0.5 to 1.25 times the ground truth values,
draw the initial TVP drift standard deviation parameters from a wide uniform prior, i.e., $\mathcal{U}(0.05,10)$, to allow for large initial variances, 
and set the discount factor (defined in Step 1) as $\delta = 0.96$.
We further assume that the filter state evolution and observation noise variances are time-invariant and specify values for each example below.
All simulations were run using MATLAB programming language (The Mathworks, Inc., Natick, MA).
In particular, we utilize MATLAB's built-in ODE solver \texttt{ode15s} in the forward propagation of the model states (see Step 2 in Section~\ref{Sec:PF}) to account for possible stiffness in the ODE systems.

\subsection{Example: 1D Advection Equation}

\begin{figure}[t!]
\centerline{\includegraphics[width=.33\linewidth]{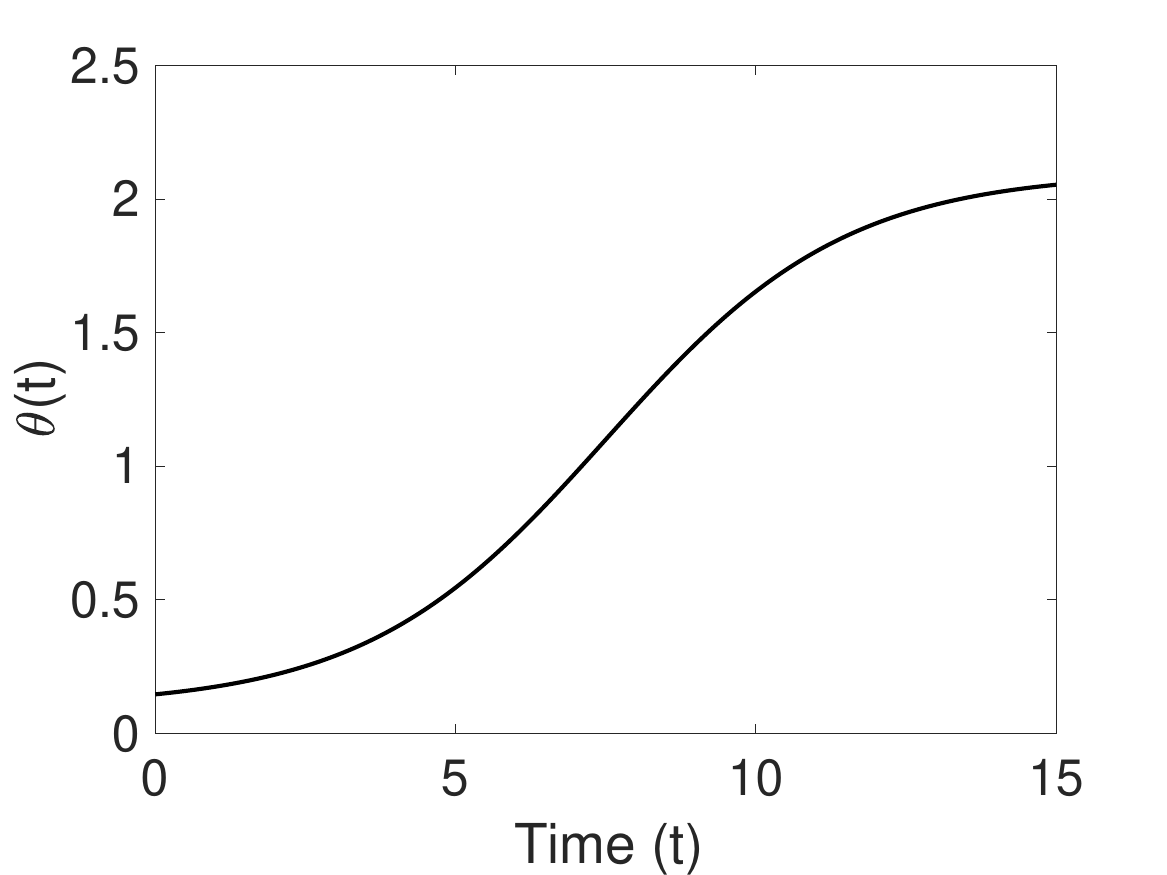} \includegraphics[width=.33\linewidth]{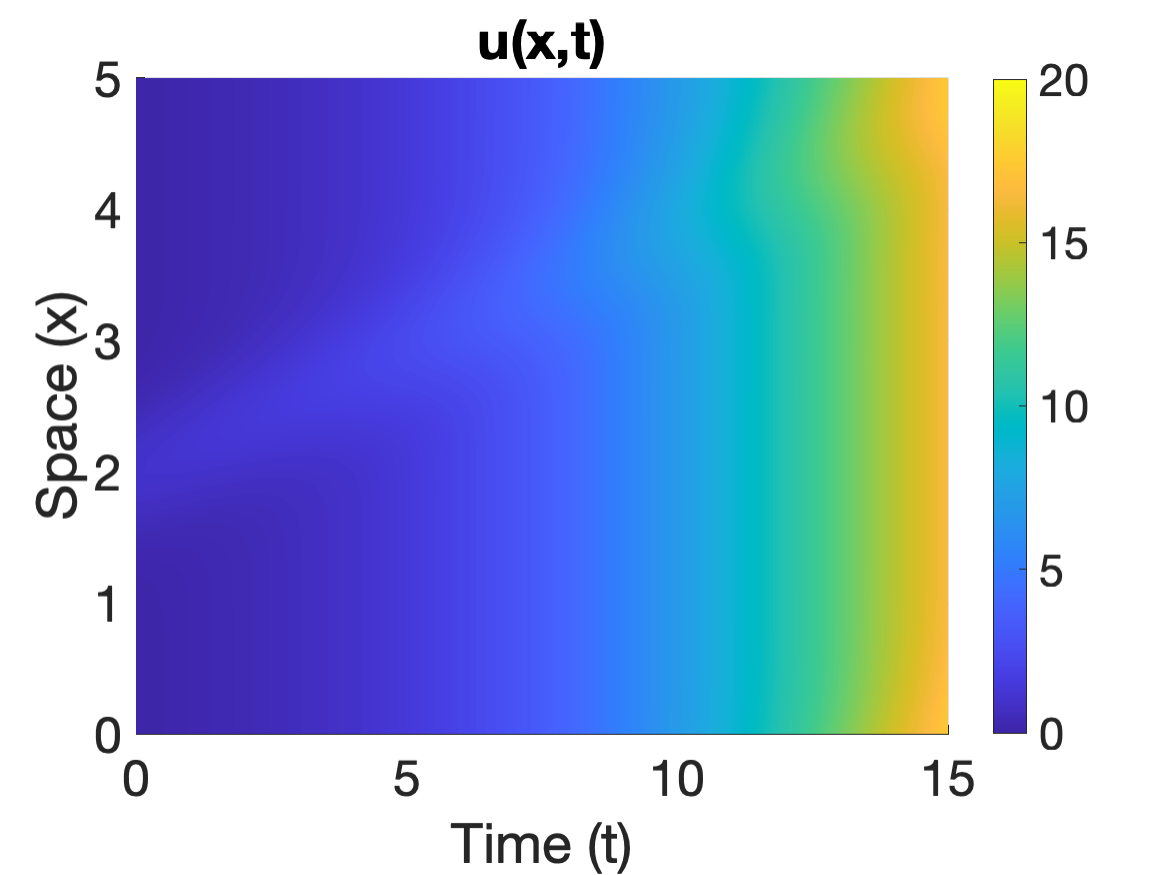}\includegraphics[width=.33\linewidth]{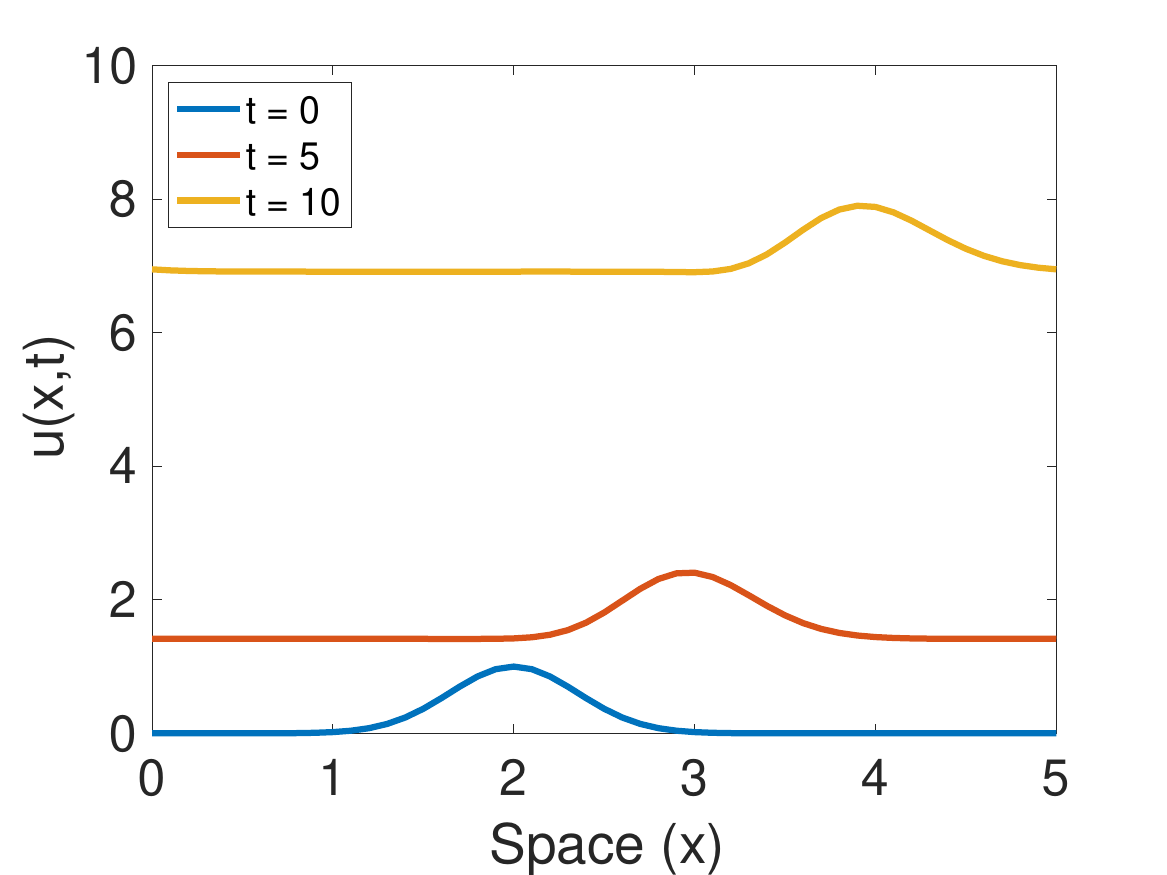}}
\caption{Simulated solution of the 1D advection equation in \eqref{Eq:1Dadv} with velocity $v=0.2$ and the logistic source parameter defined in \eqref{eq:adv_logistic_source}. From left to right: true source parameter $\theta(t)$; true solution $u(x,t)$; and solution profiles at different fixed times.}
\label{Fig:adv_sol}
\end{figure}

As a first example, consider a 1D advection equation of the form 
\begin{equation}\label{Eq:1Dadv}
\frac{\partial u}{\partial t} + v \frac{\partial u}{\partial x} = S(x,t), \quad S(x,t) = \theta(t) 
\end{equation}
where $v$ is the known advection velocity, and we model the source term $S(x,t)$ as constant over space but varying with time.
We assume a finite spatial domain $x\in[0,L]$ with $L=5$ and periodic boundary conditions such that $u(0,t) = u(L,t)$ for all $t\geq 0$.
We further assume that the initial condition $u(x,0) = \exp\left\{-(x-\mu)^2/\gamma^2 \right\}$ has a Gaussian profile with mean $\mu=2$ and standard deviation $\gamma=0.5$ and that the true source parameter is modeled as a shifted logistic function with 
\begin{equation}\label{eq:adv_logistic_source}
\theta(t) = \frac{2}{1+\exp\left\{-0.5(t-7.5)\right\}} + 0.1.
\end{equation}
The time-dependent source parameter $\theta(t)$ is our TVP of interest.
Figure~\ref{Fig:adv_sol} shows the simulated solution over the time interval [0,15] when setting $v=0.2$ as the fixed velocity.
In solving the inverse problem, we assume partial observations of $u$ at a subset of discrete spatial locations (i.e., $x = 0.1$, 0.3, \dots, 4.9, with each observed location being 0.2 spatial units apart) every 0.05 time units, corrupted by Gaussian noise.

In the offline phase, we discretize the spatial domain into $M+1$ equidistant points $x_i$, $i=0,\dots,M$, where $h = \Delta x = 0.1$ and $M=50$, and let $\mbf{u}_i(t) \approx u(x_i,t)$.
Using a central difference approximation for the spatial derivative term in \eqref{Eq:1Dadv} and 
incorporating the periodic boundary conditions such that $\mbf{u}_0(t) = \mbf{u}_M(t)$ for all $t\geq0$ leads to 
\begin{equation}
	\frac{d\mbf{u}(t)}{dt} = \mathsf{A}\mbf{u}(t) + \theta(t) \mbf{1}, 
\end{equation}
where $\mbf{u}=(\mbf{u}_1,\dots,\mbf{u}_M)^T\in\R^M$ is the vector of ODE model states, 
\begin{equation}
	\mathsf{A} = -\frac{v}{2h} \left[\begin{array}{ccccc} \phantom{-}0 & \phantom{-}1 & & & -1 \\ -1 & \phantom{-}0 & \phantom{-}1 & & \\ & \ddots & \ddots & \ddots & \\ & & -1 & \phantom{-}0 & \phantom{-}1 \\ \phantom{-}1 & & & -1 & \phantom{-}0 \end{array} \right]_{M \times M},
\end{equation}
and $\mbf{1}$ denotes the $M\times 1$ vector of all 1 entries.

\begin{figure}[t!]
\centerline{\includegraphics[width=.33\linewidth]{adv_logistic_truesol} \includegraphics[width=.33\linewidth]{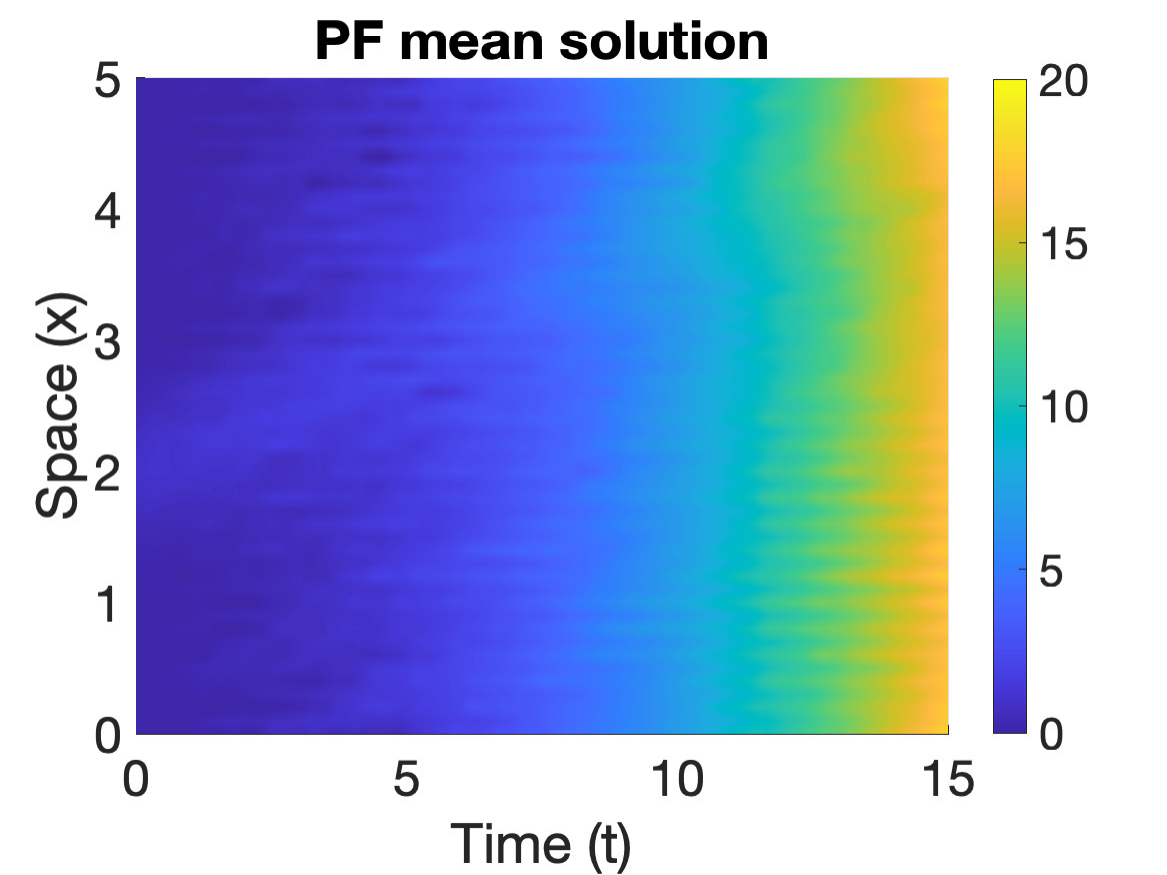}\includegraphics[width=.33\linewidth]{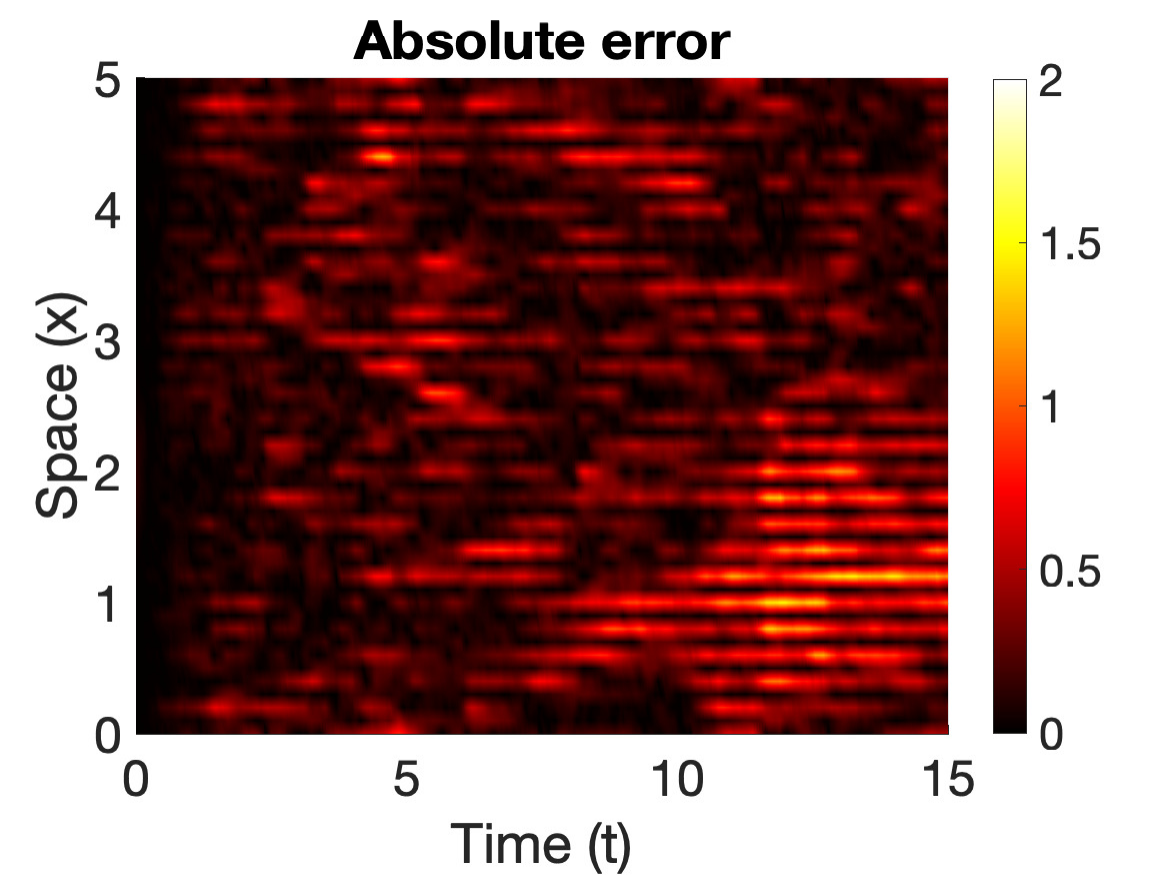}}
\caption{Results of the 1D advection example. From left to right: true solution $u(x,t)$; estimated solution using the PF mean; and absolute error between the true and estimated solutions.}
\label{Fig:adv_results_sol}
\end{figure}

\begin{figure}[t!]
\centerline{\fbox{\includegraphics[width=0.5\textwidth]{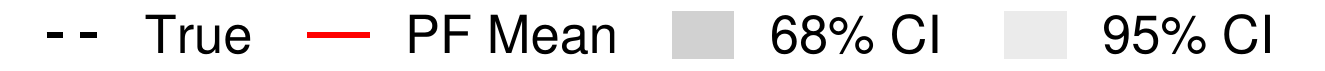} }}
\vspace{0.1cm}
\centerline{\includegraphics[width=.4\linewidth]{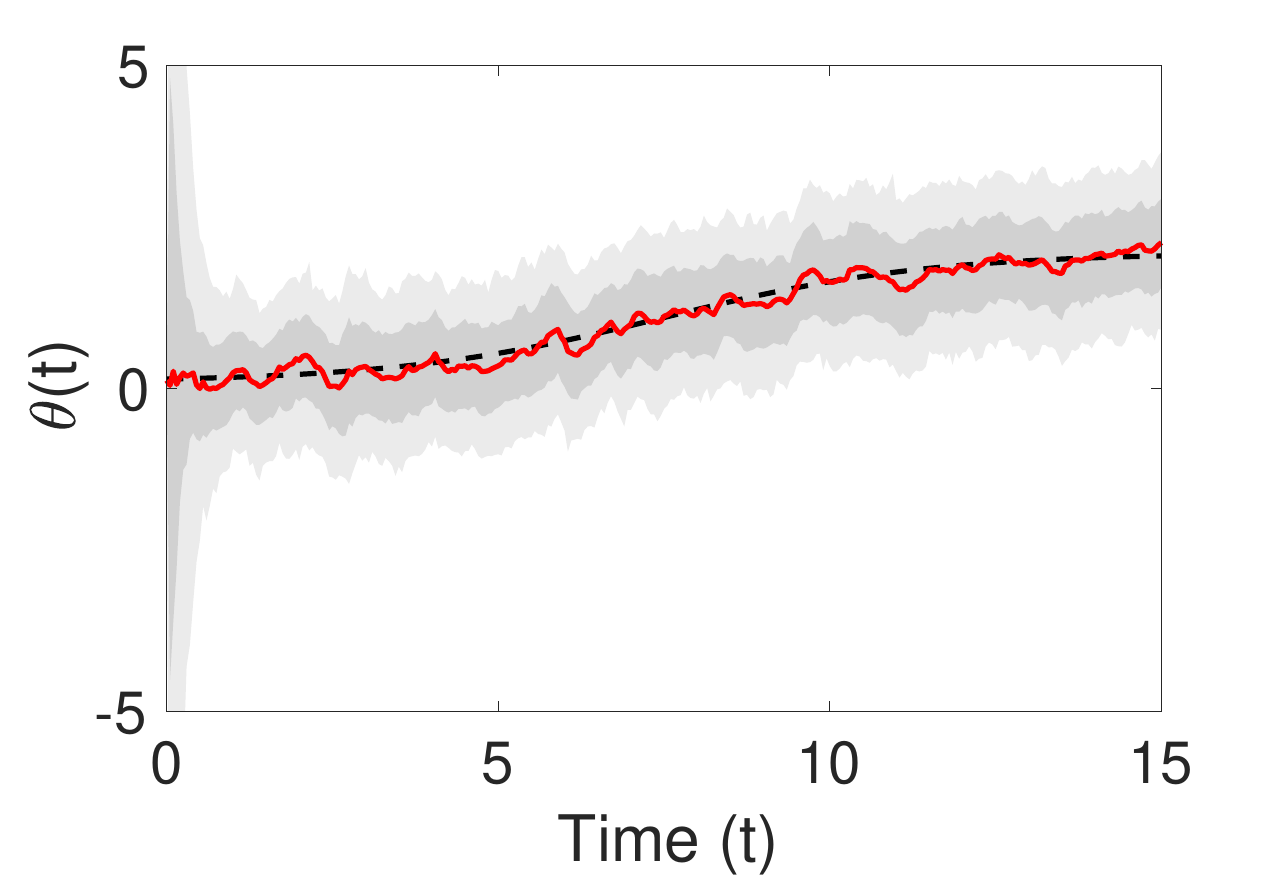} \includegraphics[width=.4\linewidth]{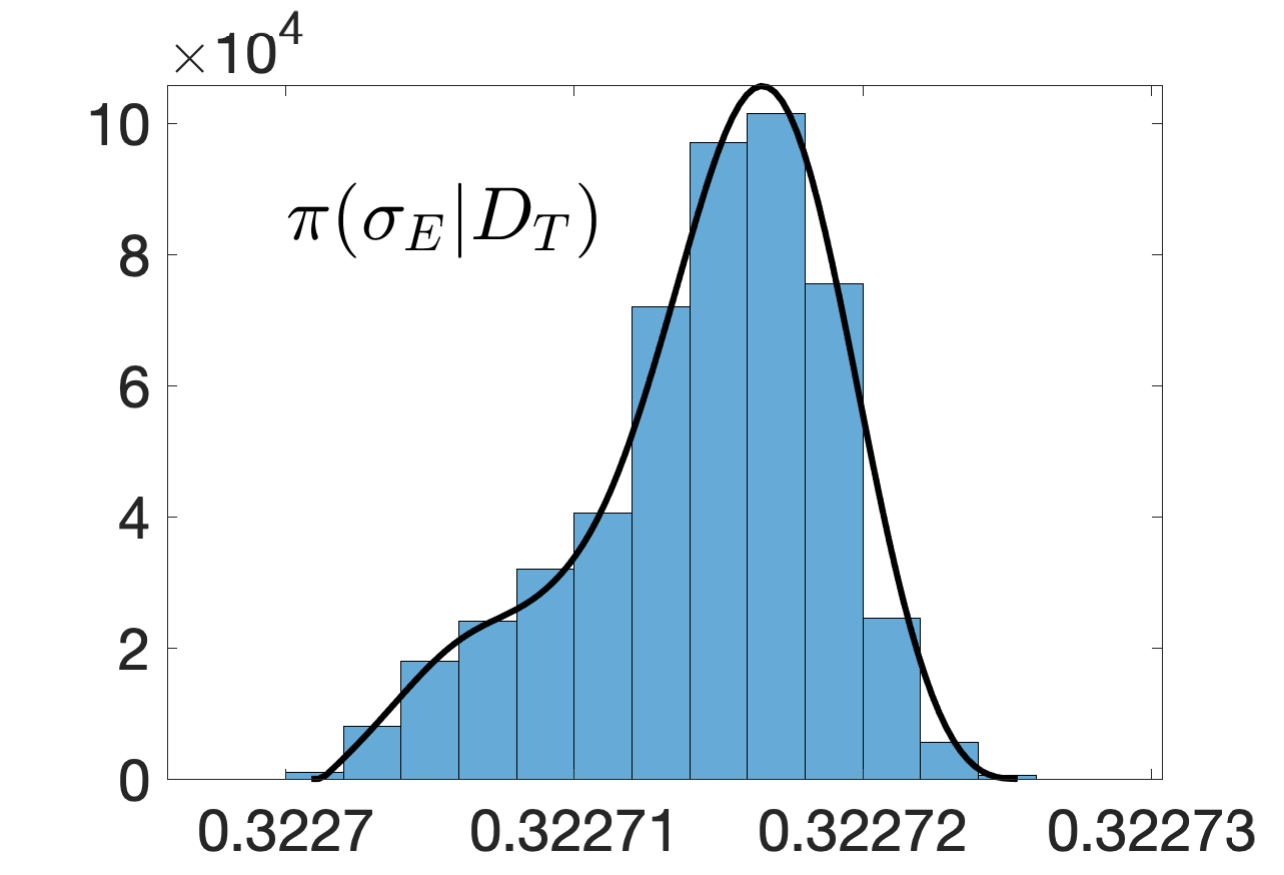}}
\centerline{\includegraphics[width=.4\linewidth]{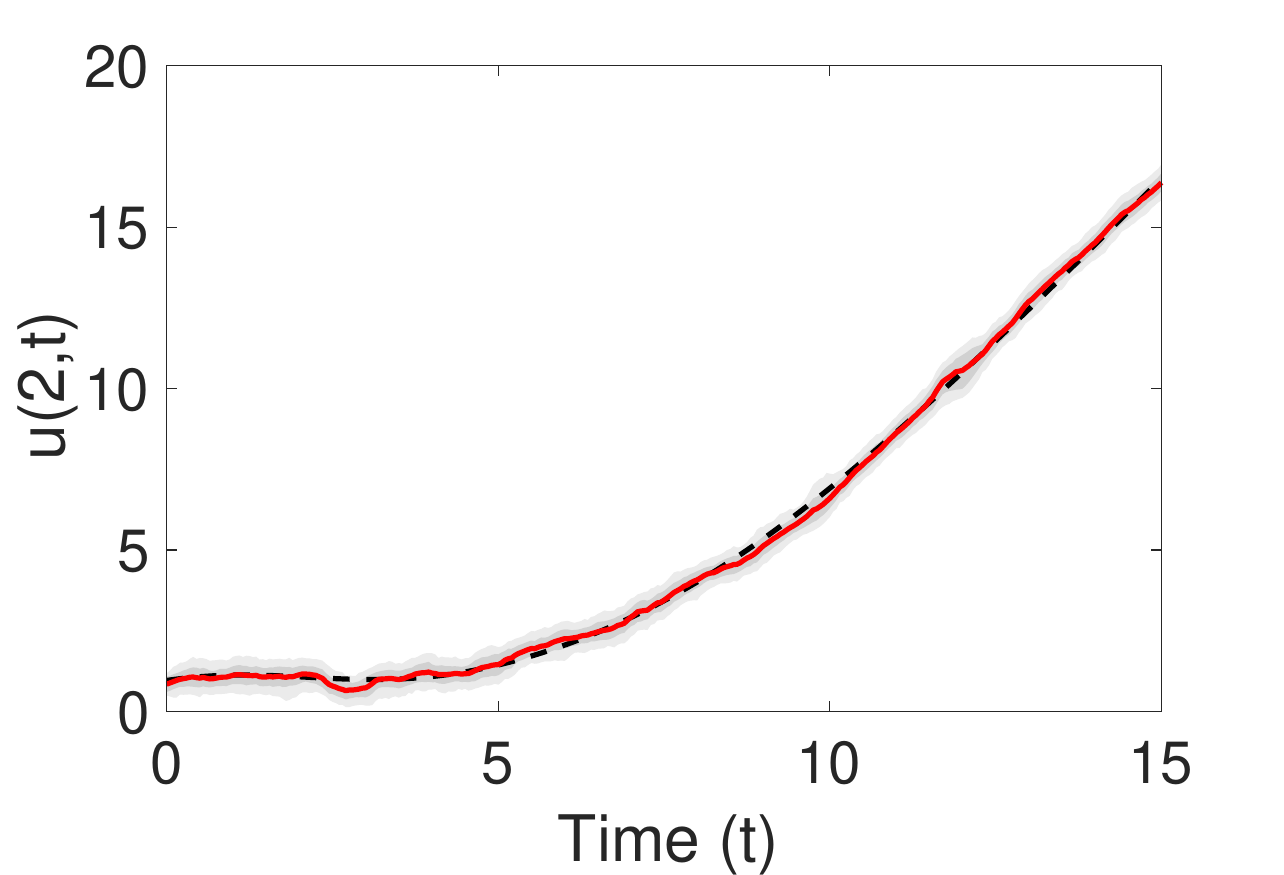} \includegraphics[width=.4\linewidth]{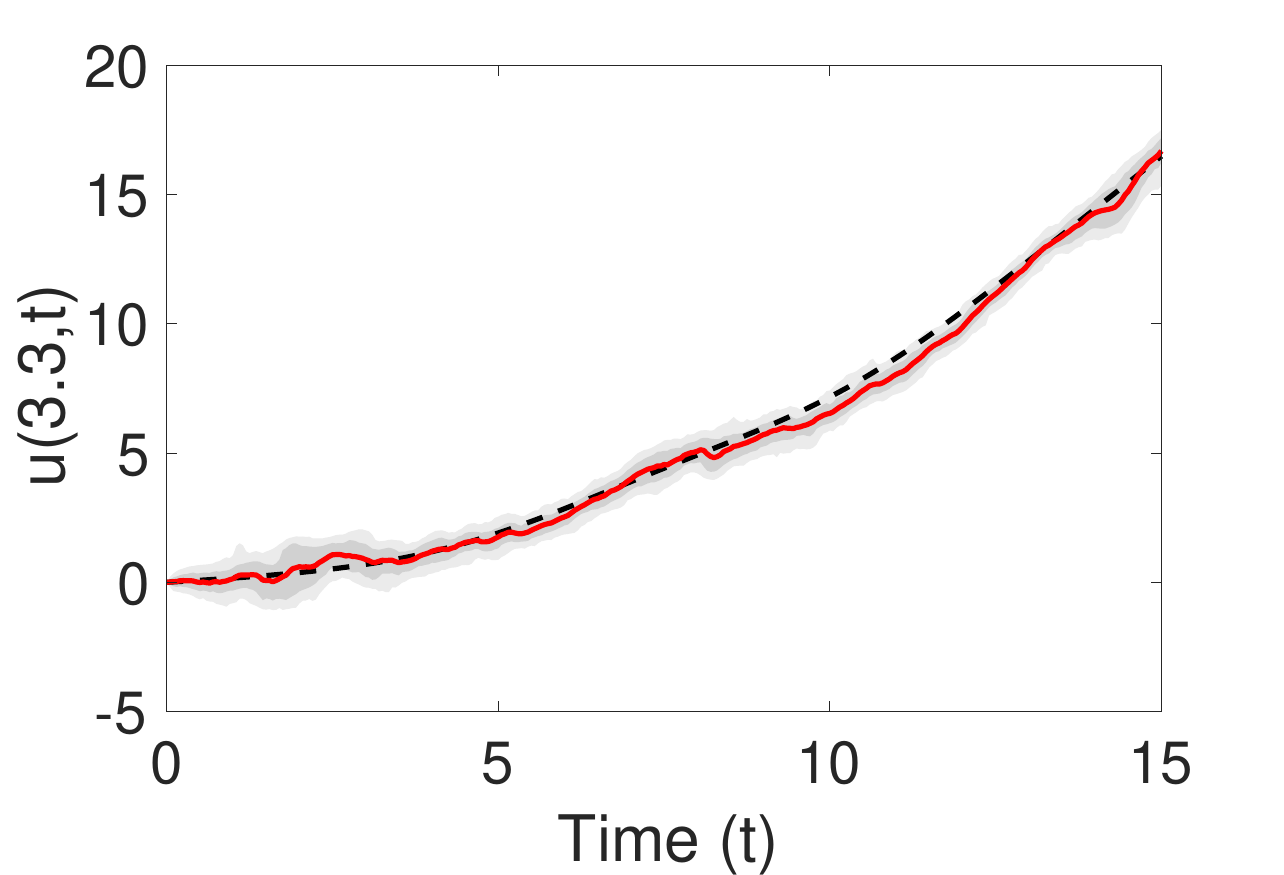}}
\caption{Results of the 1D advection example. Top row: estimated $\theta(t)$ (left) and corresponding posterior histogram of $\sigma_\mathsf{E}$ (right).  Bottom row: estimated solution trajectories at $x=2$ (left) and $x=3.3$ (right).  The true TVP and solution trajectories are shown in dashed black, the PF mean solution is shown in red, and the 68\% and 95\% Bayesian credible intervals are shown in dark and light grey, respectively.}
\label{Fig:adv_results_UQ}
\end{figure}

In the online phase, we fix the PF model innovation and observation noise standard deviations to $\sigma_\mathsf{C}=0.1$ and $\sigma_\mathsf{D}=0.75$, respectively, and proceed with the filtering algorithm as described in Section~\ref{Sec:PF}.
Figure~\ref{Fig:adv_results_sol} displays the resulting PF mean solution compared to the true solution, while Figure~\ref{Fig:adv_results_UQ} shows the corresponding estimate of $\theta(t)$ along with estimated solution trajectories at two spatial locations ($x=2$ and $x=3.3$). 
The PF mean solution in Figure~\ref{Fig:adv_results_sol} provides a fairly accurate approximation to the true solution, though there are some artifacts that could be, e.g., partially due to numerical error.
As shown in Figure~\ref{Fig:adv_results_UQ}, the filter is able to well approximate the true underlying logistic TVP, with the Bayesian credible intervals giving a measure of uncertainty around the resulting PF mean estimate.
The resulting posterior of the parameter drift coefficient is very tight around the value $\sigma_\mathsf{E} \approx 0.322$.
The filter also well tracks the solution trajectories at the two spatial locations considered, where $x=3.3$ was an observed state while $x=2$ was not observed.
In estimating the solution trajectories in these locations, we note that the credible intervals are fairly tight around the PF mean but capture the true solutions in each case.

\subsection{Example: 1D Heat Equation}

\begin{figure}[t!]
\centerline{\includegraphics[width=.33\linewidth]{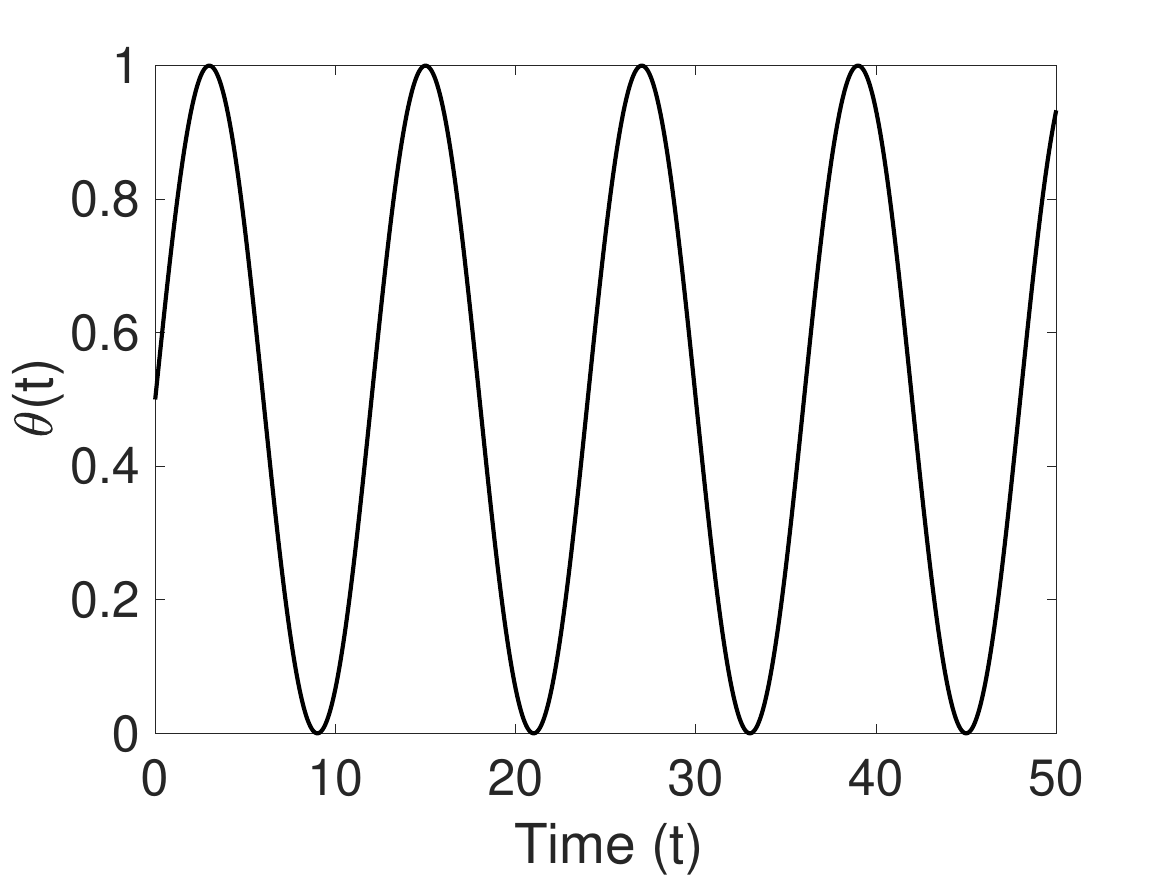} \includegraphics[width=.33\linewidth]{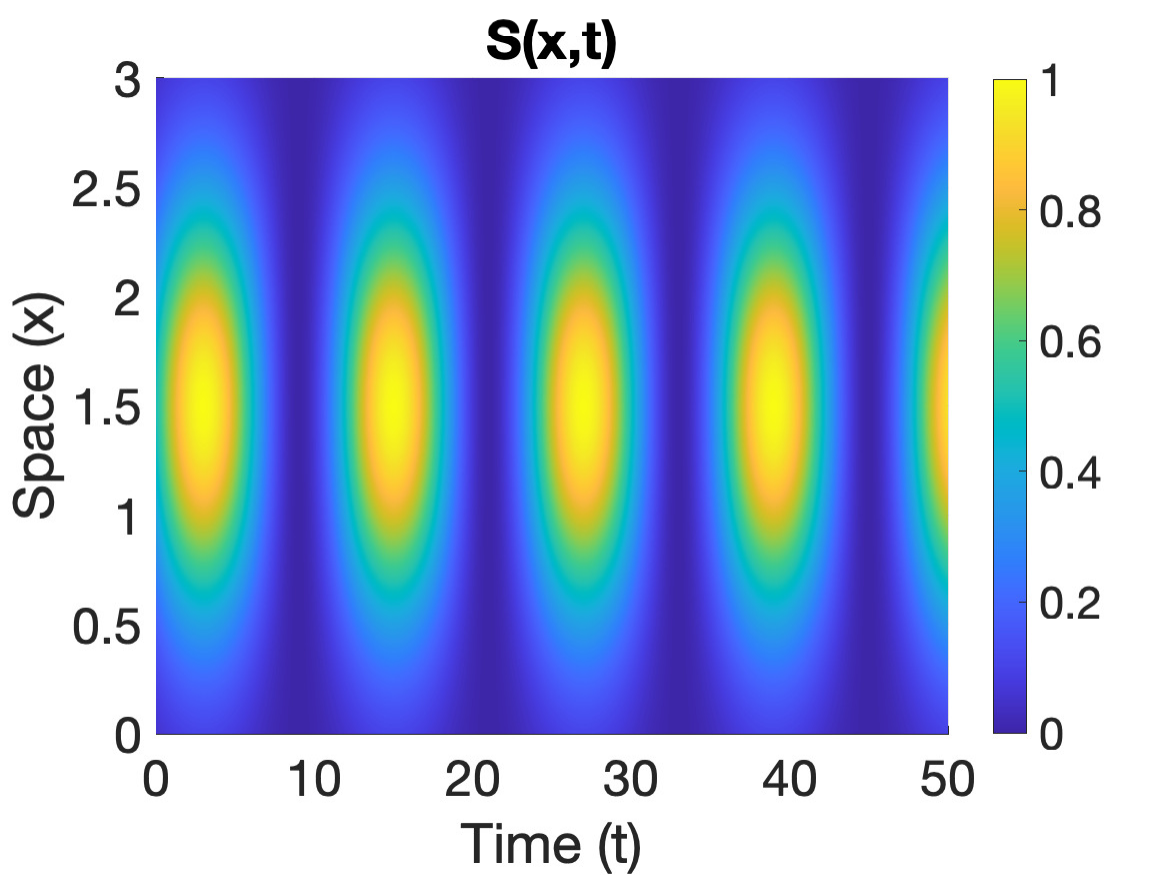}\includegraphics[width=.33\linewidth]{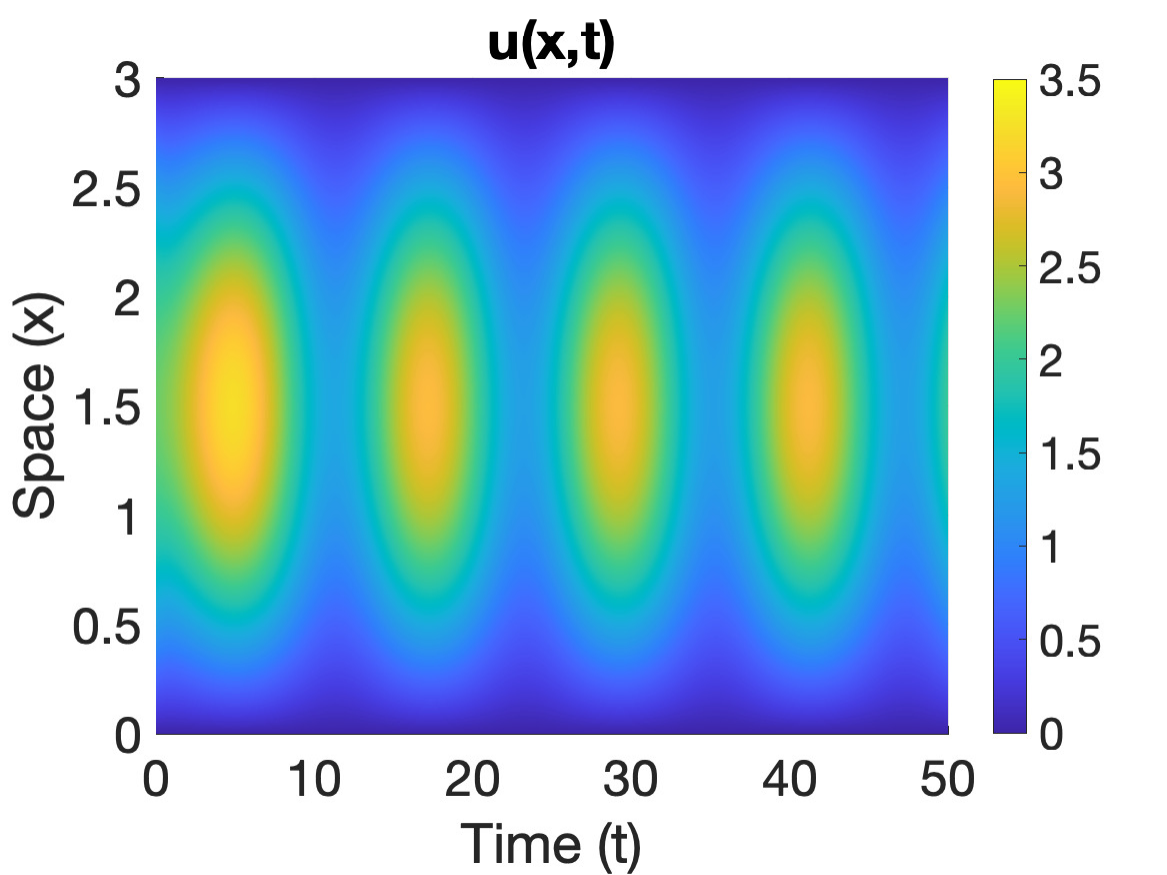}}
\caption{Simulated solution of the 1D heat equation in \eqref{eq:1Dheat} with diffusion coefficient $\alpha=0.2$ and Gaussian source with mean $\mu=1.5$, variance $\gamma^2=1$, and sinusoidal amplitude parameter $\theta(t) = 0.5\sin(\frac{\pi}{6} t)+0.5$. From left to right: true source parameter $\theta(t)$; true source $S(x,t)$; and true solution $u(x,t)$.}
\label{Fig:heat_sol}
\end{figure}

As a second example, consider a 1D heat equation of the form 
\begin{equation} \label{eq:1Dheat}
\frac{\partial u}{\partial t} = \alpha \frac{\partial^2 u}{\partial x^2} + S(x,t), \quad S(x,t) = \theta(t)\exp{\big\{-(x-\mu)^2/\gamma^2 \big\}}
\end{equation}
where we assume $x\in [0,3]$, $t\in[0,50]$, and $S(x,t)$ is a Gaussian source term with time-dependent amplitude parameter $\theta(t)$.
Note that we can write the source term as $S(x,t) = \theta(t)s(x)$, separating the unknown TVP from the spatially-dependent term $s(x)$. 
We further assume Dirichlet boundary conditions $u(0,t) = u(3,t) = 0$ and initial condition $u(x,0) = 3x - x^2$.
A similar model with different parameterization is considered in \cite{Qin2021}.
Figure~\ref{Fig:heat_sol} shows the simulated solution over the time interval [0,50] when $\alpha=0.2$ and the true source term is parameterized with $\theta(t) = 0.5\sin(\frac{\pi}{6} t)+0.5$, $\mu = 1.5$, and $\gamma=1$.
In solving the inverse problem, we partially observe $u$ at 8 equidistant spatial locations (i.e., at $x = 0.1$, 0.5, \dots, 2.9, each observed trajectory being 0.4 spatial units apart) every 0.1 time units, with observations corrupted by Gaussian noise.

In the offline phase, we implement a spatial discretization $x_i$, $i=0,\dots, M$, over the interval [0,3] with step size $h=0.1$ and $M=30$, and we again denote $\mbf{u}_i(t) \approx u(x_i,t)$.
Applying the Dirichlet boundary conditions $\mbf{u}_0(t) = \mbf{u}_M(t) = 0$ and a central difference approximation to the second-order spatial derivative in \eqref{eq:1Dheat} yields the following ODE system:
\begin{equation}
    \frac{d\mbf{u}(t)}{dt} = \mathsf{B}\mbf{u}(t) + \theta(t) \mbf{s}, 
\end{equation}
where $\mbf{u}(t) = \left(\mbf{u}_1(t), \dots, \mbf{u}_{M-1}(t) \right)^T \in\R^{M-1}$, 
\begin{equation}
    \mathsf{B} = \frac{\alpha}{h^2} \left[\begin{array}{ccccc} -2 & \phantom{-}1 & & & \\ \phantom{-}1 & -2 & \phantom{-}1 & & \\ & \ddots & \ddots & \ddots & \\ & & \phantom{-}1 & -2 & \phantom{-}1 \\ & & & \phantom{-}1 & -2 \end{array} \right]_{(M-1)\times (M-1)} \quad \text{and} \quad \mbf{s} = \left[\begin{array}{c} s(x_1) \\ \\ \vdots \\ \\ s(x_{M-1}) \end{array} \right]_{(M-1)\times 1}.
\end{equation}

\begin{figure}[t!]
\centerline{\includegraphics[width=.33\linewidth]{heat_sine_truesol} \includegraphics[width=.33\linewidth]{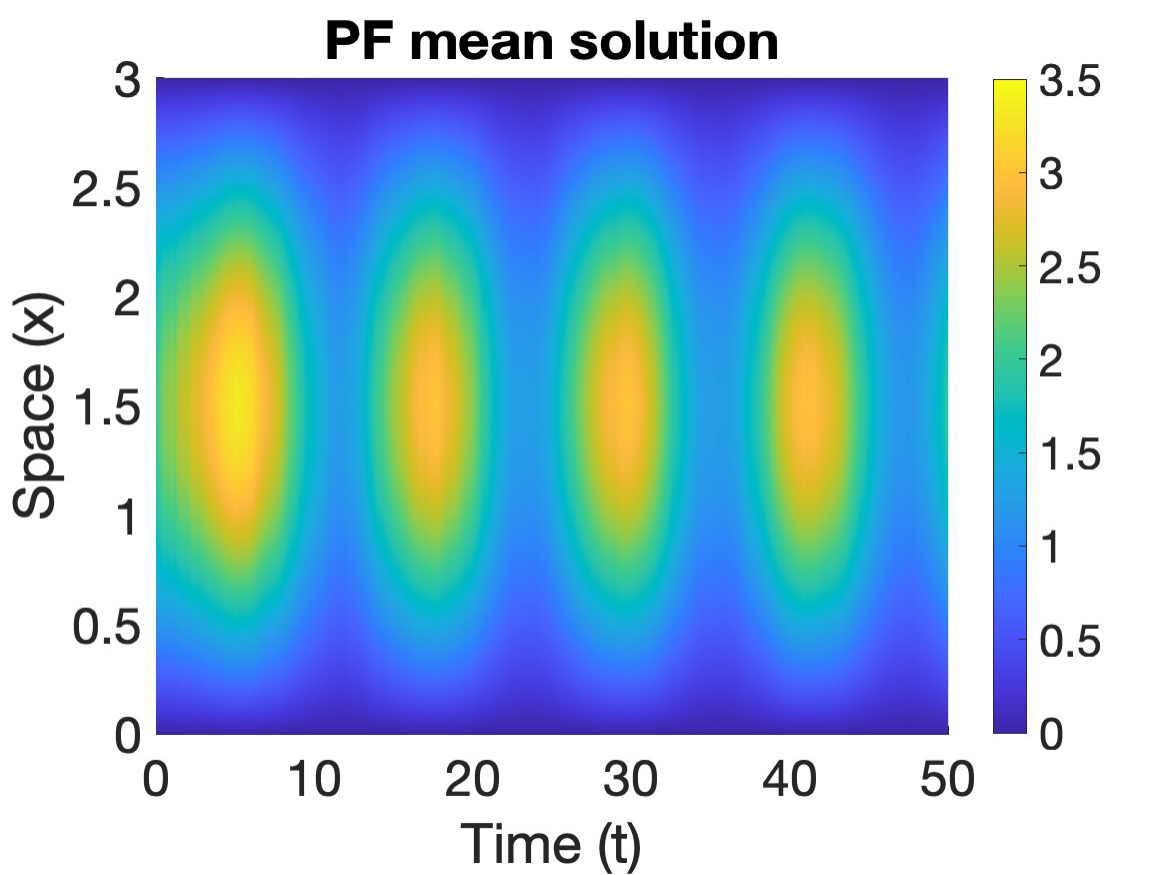}\includegraphics[width=.33\linewidth]{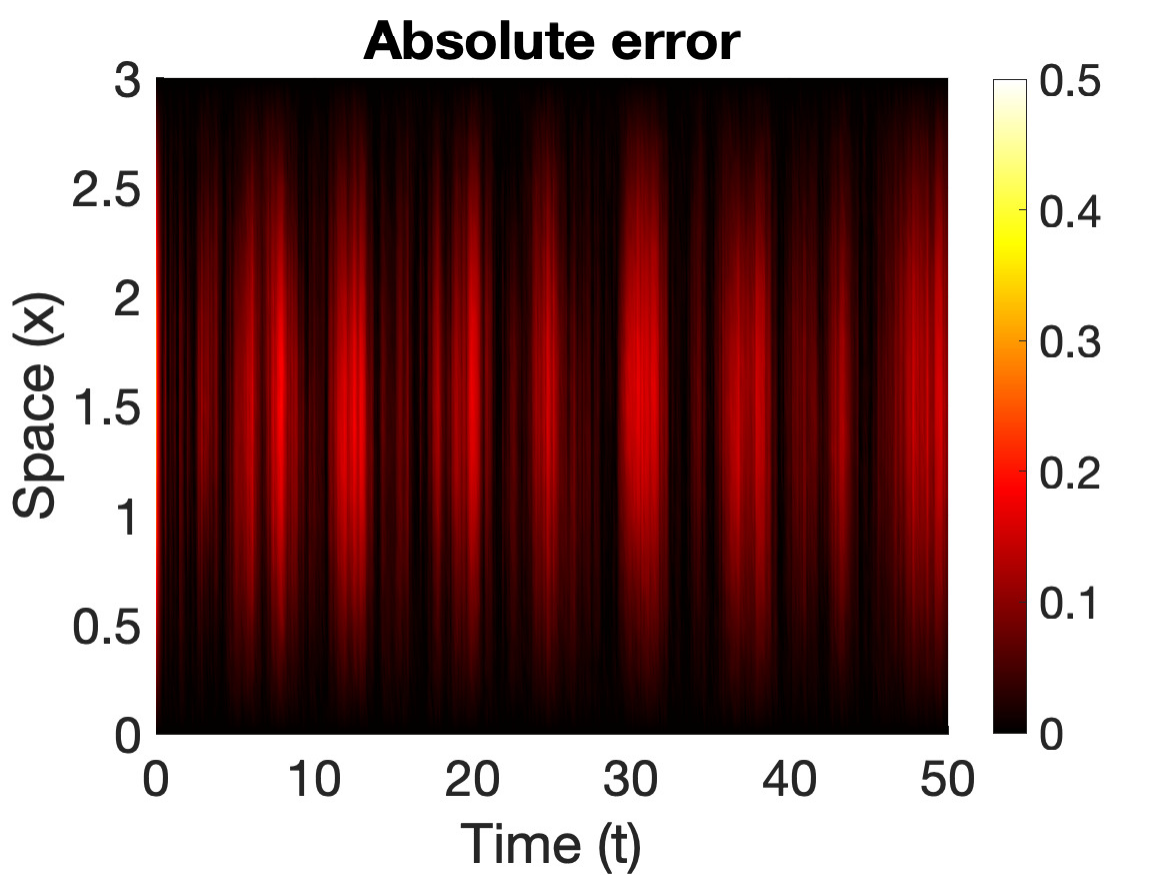}}
\caption{Results of the 1D heat example. From left to right: true solution $u(x,t)$; estimated solution using the PF mean; and absolute error between the true and estimated solutions.}
\label{Fig:heat_results_sol}
\end{figure}

\begin{figure}[t!]
\centerline{\fbox{\includegraphics[width=0.5\textwidth]{results_legend_fixedsigma} }}
\vspace{0.1cm}
\centerline{\includegraphics[width=.4\linewidth]{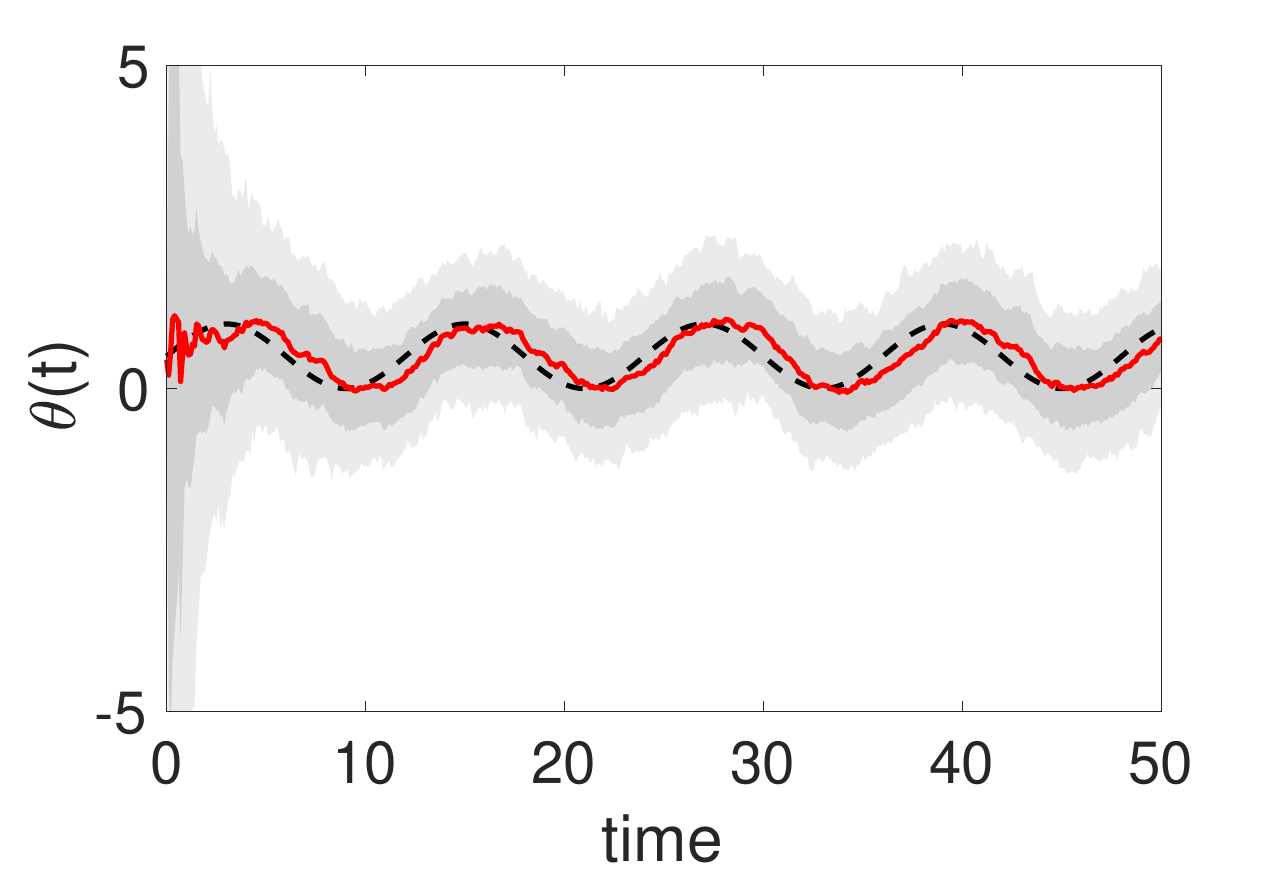} \includegraphics[width=.4\linewidth]{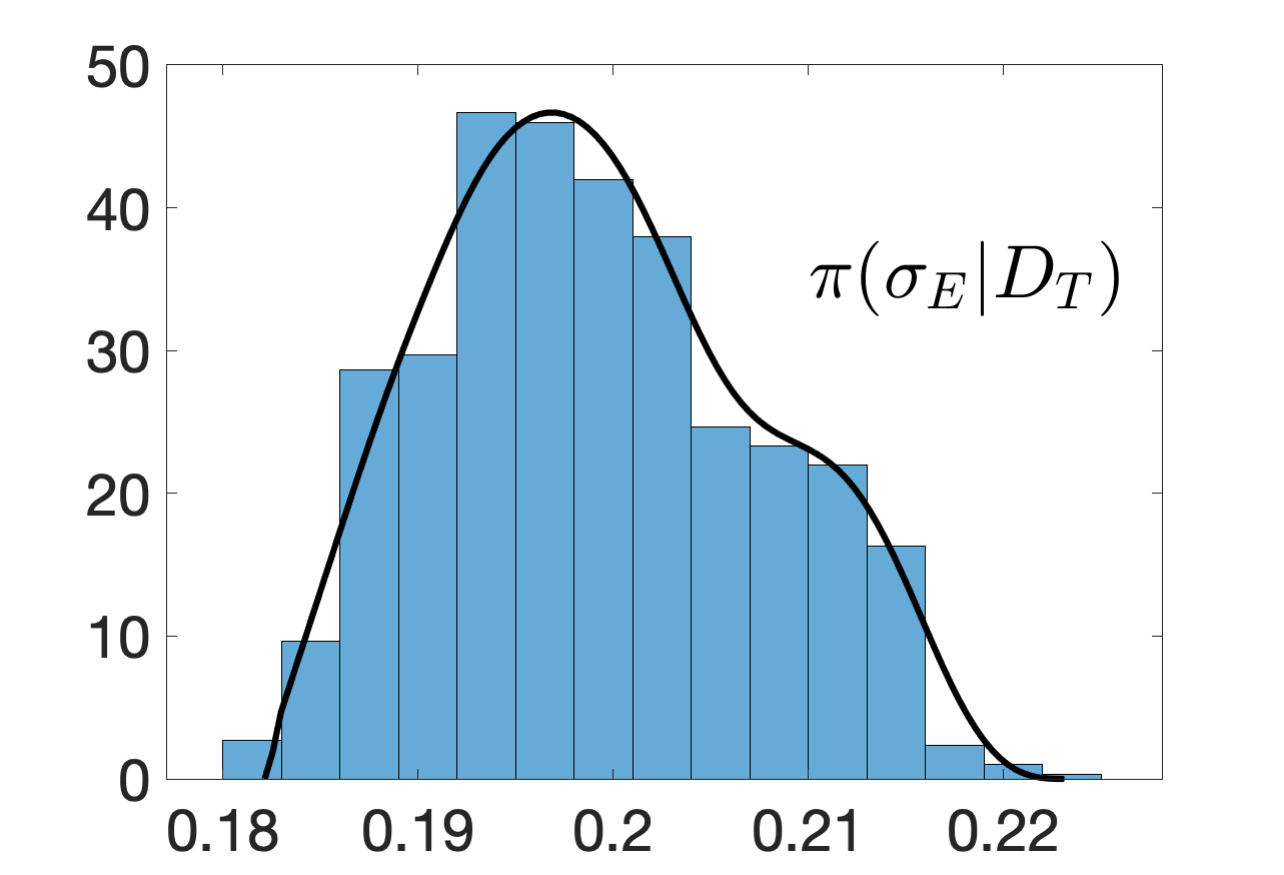}}
\centerline{\includegraphics[width=.4\linewidth]{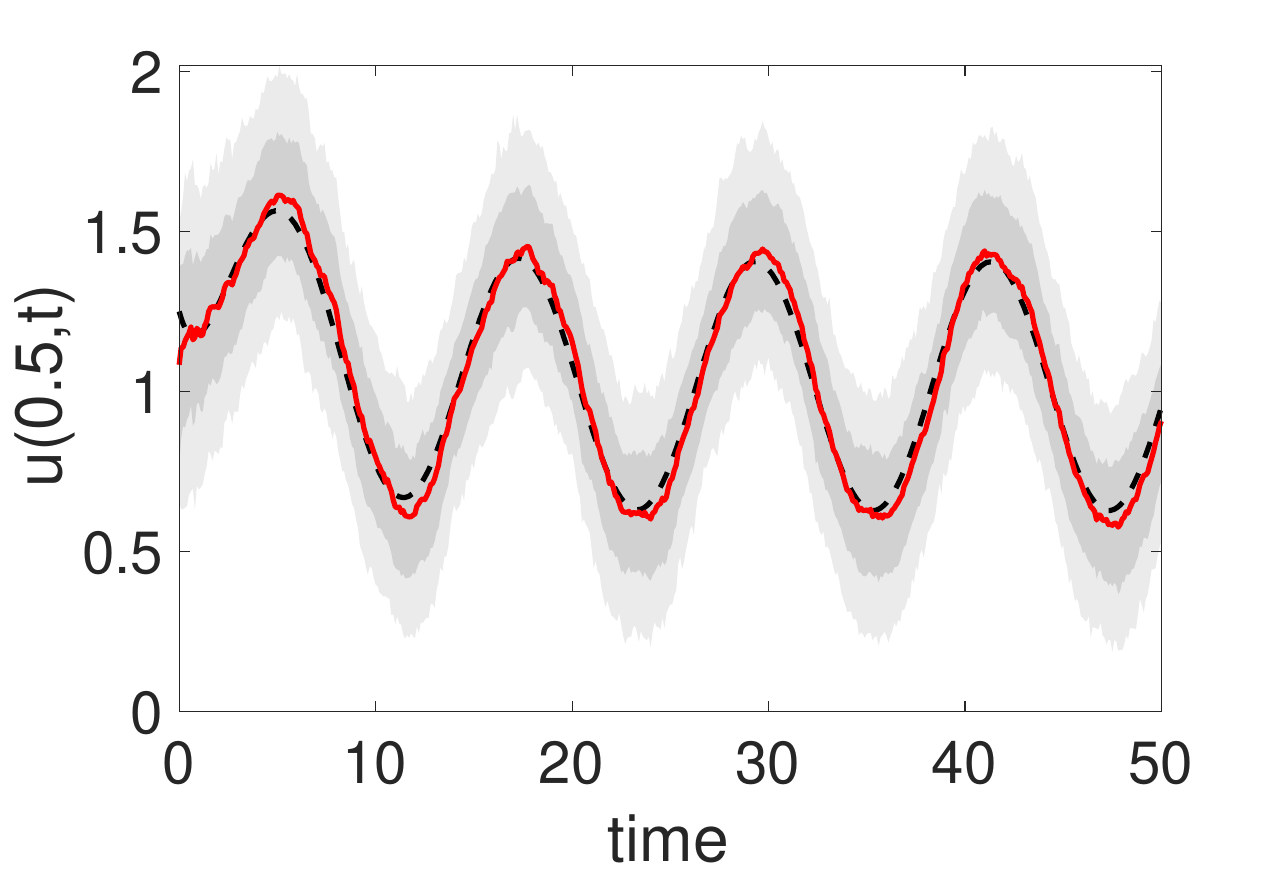} \includegraphics[width=.4\linewidth]{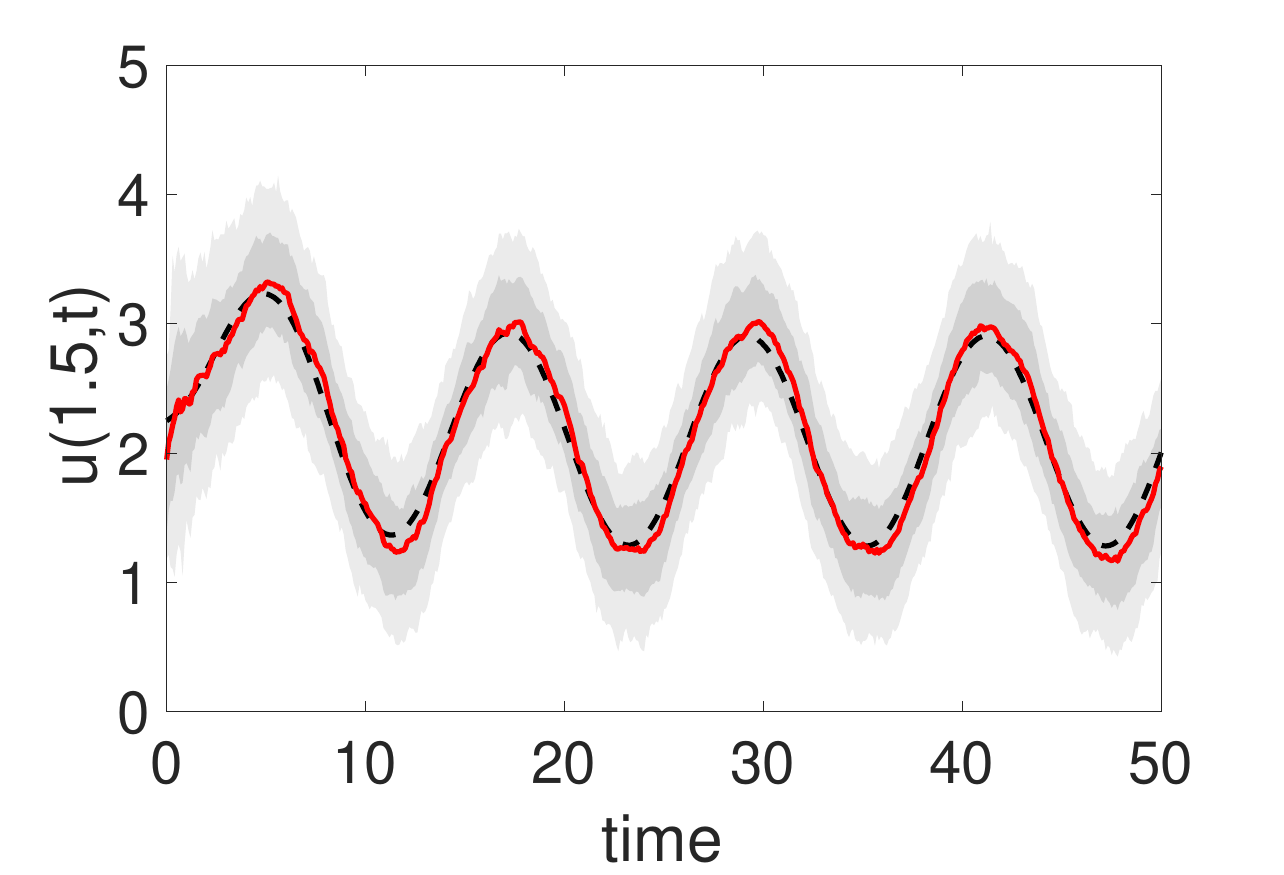}}
\caption{Results of the 1D heat example. Top row: estimated $\theta(t)$ (left) and corresponding posterior histogram of $\sigma_\mathsf{E}$ (right).  Bottom row: estimated solution trajectories at $x=0.5$ (left) and $x=1.5$ (right).  The true TVP and solution trajectories are shown in dashed black, the PF mean solution is shown in red, and the 68\% and 95\% Bayesian credible intervals are shown in dark and light grey, respectively.}
\label{Fig:heat_results_UQ}
\end{figure}

For the online filtering phase, we set $\sigma_\mathsf{C}=0.1$ and $\sigma_\mathsf{D}=1.5$.
Figure~\ref{Fig:heat_results_sol} shows the resulting mean PF solution compared to the true solution, and Figure~\ref{Fig:heat_results_UQ} gives the resulting TVP estimate and estimated solution trajectories at spatial locations $x=0.5$ and $x=1.5$.
As seen in Figure~\ref{Fig:heat_results_sol}, the PF mean solution provides a reasonably accurate estimate of the true solution, with less error than in the previous example. 
The filter is similarly able to well track both the true underlying TVP and solution trajectories at the two spatial locations considered (here, $x=0.5$ was an observed state but $x=1.5$ was not observed), and the credible intervals fully capture the true values in each case.

\section{Discussion and Conclusions}
\label{Sec:Discussion}

On the 1D examples shown in this work, the proposed estimation procedure proves robust in tracking unknown TVPs in the source terms of advection-diffusion-type PDEs from partial, noisy observations of the solution at discrete spatial locations and times.  
While the true underlying TVPs are closely tracked by the PF mean in each example, we see less error in reconstructing the PDE solution $u(x,t)$ from the PF mean in the heat equation example as opposed to the advection equation example, which is typically considered to be more difficult to solve numerically.
In both examples, the PF mean well approximates the solution trajectories at both observed and unobserved spatial locations, with the true solutions contained within the credible interval uncertainty bounds.

While the numerical results in this work focused on 1D examples, we note that the proposed framework accommodates extension to higher spatial dimensions (e.g., 2D and 3D PDE models), affecting the offline discretization phase and resulting in a larger ODE system to solve during the online phase, potentially causing a significant increase in computing time.
To address this bottleneck, future work will consider efficient ways to implement the forward propagation step in the PF, e.g., using vectorized computation with multistep solvers, as studied in \cite{Arnold2015}.  
A key element for numerical stability is to select the spatial discretization step size and the integration time step size in order to ensure that the Courant-Friedrichs-Lewy (CFL) condition is met, which becomes more challenging if parameters such as the advection velocity or diffusion coefficient are also time-varying.
Future work will address extension of this technique to consider estimating TVPs of these types as well as spatially-dependent parameters.


\section*{ORCID iD}
Andrea Arnold: \url{https://orcid.org/0000-0003-3003-882X}



\bibliography{AA_refs}{}

\begin{thebibliography}{10}

\bibitem{Wang2020}
D.~Wang, K.~Liu, and X.~Zhang.
\newblock Spatiotemporal thermal field modeling using partial differential
  equations with time-varying parameters.
\newblock {\em IEEE Transactions on Automation Science and Engineering},
  17(2):646--657, 2020.

\bibitem{Rodrigo2006}
M.~R. Rodrigo and R.~S. Mamon.
\newblock An alternative approach to solving the {Black-Scholes} equation with
  time-varying parameters.
\newblock {\em Applied Mathematics Letters}, 19:398--402, 2006.

\bibitem{Selvadurai2004}
A.~P.~S. Selvadurai.
\newblock On the advective-diffusive transport in porous media in the presence
  of time-dependent velocities.
\newblock {\em Geophysical Research Letters}, 31:L13505, 2004.

\bibitem{ArnoldFichera2022}
A.~Arnold and L.~Fichera.
\newblock Identification of tissue optical properties during thermal
  laser-tissue interactions: An ensemble {K}alman filter-based approach.
\newblock {\em International Journal for Numerical Methods in Biomedical
  Engineering}, 38(4):e3574, 2022.

\bibitem{Arnold2023}
A.~Arnold.
\newblock When artificial parameter evolution gets real: particle filtering for
  time-varying parameter estimation in deterministic dynamical systems.
\newblock {\em Inverse Problems}, 39(1):014002, 2023.

\bibitem{Jamili2021}
E.~Jamili and V.~Dua.
\newblock Parameter estimation of partial differential equations using
  artificial neural network.
\newblock {\em Computers and Chemical Engineering}, 147:107221, 2021.

\bibitem{Muller2002}
T.~G. Muller and J.~Timmer.
\newblock Fitting parameters in partial differential equations from partially
  observed noisy data.
\newblock {\em Physica D}, 171:1--7, 2002.

\bibitem{Carvalho2015}
E.~P. Carvalho, J.~Martinez, J.~M. Martinez, and F.~Pisnitchenko.
\newblock On optimization strategies for parameter estimation in models
  governed by partial differential equations.
\newblock {\em Mathematics and Computers in Simulation}, 114:14--24, 2015.

\bibitem{Guo2009}
L.~Z. Guo, S.~A. Billings, and D.~Coca.
\newblock Consistent recursive parameter estimation of partial differential
  equation models.
\newblock {\em International Journal of Control}, 82(10):1946--1954, 2009.

\bibitem{Raissi2017}
M.~Raissi, P.~Perdikaris, and G.~E. Karniadakis.
\newblock Machine learning of linear differential eequations using {Gaussian}
  processes.
\newblock {\em Journal of Computational Physics}, 348:683--693, 2017.

\bibitem{Rai2019}
P.~K. Rai and S.~Tripathi.
\newblock Gaussian process for estimating parameters of partial differential
  equations and its application to the {Richards} equation.
\newblock {\em Stochastic Environmental Research and Risk Assessment},
  33:1629--1649, 2019.

\bibitem{Xun2013}
X.~Xun, J.~Cao, B.~Mallick, A.~Maity, and R.~J. Carroll.
\newblock Parameter estimation of partial differential equation models.
\newblock {\em Journal of the American Statistical Association},
  108(503):1009--1020, 2013.

\bibitem{Abdolee2014}
R.~Abdolee, B.~Champagne, and A.~H. Sayed.
\newblock Estimation of space-time varying parameters using a diffusion {LMS}
  algorithm.
\newblock {\em IEEE Transactions on Signal Processing}, 62(2):403--418, 2014.

\bibitem{Kramer2013}
S.~Kramer and E.~M. Bollt.
\newblock Spatially dependent parameter estimation and nonlinear data
  assimilation by autosynchronization of a system of partial differential
  equations.
\newblock {\em Chaos}, 23:033101, 2013.

\bibitem{Young2012}
J.~Young and D.~Ridzal.
\newblock An application of random projection to parameter estimation in
  partial differential equations.
\newblock {\em SIAM Journal on Scientific Computing}, 34(4):A2344--A2365, 2012.

\bibitem{Beskos2015}
A.~Beskos, A.~Jasra, E.~A. Muzaffer, and A.~M. Stuart.
\newblock {Sequential Monte Carlo} methods for {Bayesian} elliptic inverse
  problems.
\newblock {\em Statistics and Computing}, 25:727--737, 2015.

\bibitem{Kantas2014}
N.~Kantas, A.~Beskos, and A.~Jasra.
\newblock {Sequential Monte Carlo} methods for high-dimensional inverse
  problems: a case study for the {Navier-Stokes} equations.
\newblock {\em SIAM/ASA Journal on Uncertainty Quantification}, 2:464--489,
  2014.

\bibitem{Zhang2017}
X.~Zhang, J.~Cao, and R.~J. Carroll.
\newblock Estimating varying coefficients for partial differential equation
  models.
\newblock {\em Biometrics}, 73(3):949--959, 2017.

\bibitem{Qin2021}
T.~Qin, Z.~Chen, J.~D. Jakeman, and D.~Xiu.
\newblock Data-driven learning of nonautonomous systems.
\newblock {\em SIAM Journal on Scientific Computing}, 43(3):A1607--A1624, 2021.

\bibitem{Lin2025}
Y.~Lin, H.~Huang, and X.~Zhang.
\newblock Source term estimation of a time-varying source around a building
  based on {Bayesian} inference and unsteady adjoint equations.
\newblock {\em Building and Environment}, 267:112251, 2025.

\bibitem{Beard2000}
D.~A. Beard and J.~B. Bassingthwaighte.
\newblock Advection and diffusion of substances in biological tissues with
  complex vascular networks.
\newblock {\em Annals of Biomedical Engineering}, 28:253--268, 2000.

\bibitem{Luce2013}
C.~H. Luce, D.~Tonina, F.~Gariglio, and R.~Applebee.
\newblock Solutions for the diurnally forced advection-diffusion equation to
  estimate bulk fluid velocity and diffusivity in streambeds from temperature
  time series.
\newblock {\em Water Resources Research}, 49(1):488--506, 2013.

\bibitem{Koch1987}
D.~L. Koch and J.~F. Brady.
\newblock A non-local description of advection-diffusion with application to
  dispersion in porous media.
\newblock {\em Journal of Fluid Mechanics}, 180:387--403, 1987.

\bibitem{Pacheco2025}
N.~E. Pacheco, K.~Zhang, A.~S. Reyes, C.~J. Pacheco, L.~Burstein, and
  L.~Fichera.
\newblock Towards a physics engine to simulate robotic laser surgery: Finite
  element modeling of thermal laser-tissue interactions.
\newblock In {\em Proceedings of the 2025 International Symposium on Medical
  Robotics}, pages 129--135, Atlanta, GA, 2025.

\bibitem{LiuWest2001}
J.~Liu and M.~West.
\newblock Combined parameter and state estimation in simulation-based
  filtering.
\newblock In A.~Doucet, N.~de~Freitas, and N.~Gordon, editors, {\em Sequential
  Monte Carlo Methods in Practice}, pages 197--223. Springer, New York, 2001.

\bibitem{West1993a}
M.~West.
\newblock Approximating posterior distributions by mixtures.
\newblock {\em Journal of the Royal Statistical Society}, 55:409--422, 1993.

\bibitem{West1993b}
M.~West.
\newblock Mixture models, {M}onte {C}arlo, {B}ayesian updating and dynamic
  models.
\newblock In J.~H. Newton, editor, {\em Computing Science and Statistics:
  Proceedings of the 24th Symposium on the Interface}, pages 325--333, Fairfax
  Station, VA, 1993. Interface Foundation of North America.

\bibitem{Arnold2015}
A.~Arnold, D.~Calvetti, and E.~Somersalo.
\newblock Vectorized and parallel particle filter {SMC} parameter estimation
  for stiff {ODEs}.
\newblock In M.~de~Leon, W.~Feng, Z.~Feng, X.~Lu, J.~M. Martell, J.~Parcet,
  D.~Peralta-Salas, and W.~Ruan, editors, {\em Dynamical Systems and
  Differential Equations, AIMS Proceedings 2015}, pages 75--84, Madrid, Spain,
  2015.

\end{thebibliography}

\end{document}